\documentclass[journal,draftcls,onecolumn,english]{IEEEtran}
\usepackage{amsmath,amssymb,amsfonts}
\usepackage{graphicx}
\usepackage{subfigure}
\usepackage{textcomp}
\usepackage{multirow}
\usepackage{blindtext}
\usepackage{algorithm,algpseudocode}

\def\BibTeX{{\rm B\kern-.05em{\sc i\kern-.025em b}\kern-.08em
    T\kern-.1667em\lower.7ex\hbox{E}\kern-.125emX}}

\begin{document}

\title{INRISCO: INcident monitoRing In Smart COmmunities}

\author{\IEEEauthorblockN{M\'onica Aguilar Igartua}\\
\IEEEauthorblockA{Dept. of Network Engineering. Universitat Polit\`ecnica de Catalunya. email: monica.aguilar@upc.edu}\\
\and
\IEEEauthorblockN{Florina Almenares}\\
\IEEEauthorblockA{Dept. of Telematics Engineering. Universidad Carlos III de Madrid. email: florina@it.uc3m.es}\\
\and
\IEEEauthorblockN{Rebeca P. D\'iaz Redondo}
\IEEEauthorblockA{lanTTic Research Center. Universidade de Vigo. email: rebeca@det.uvigo.es}\\
\and
\IEEEauthorblockN{Manuela I. Mart\'in}\\
\IEEEauthorblockA{Centro Tecnolóxico de Telecomunicacións de Galicia (GRADIANT). email: manuelaimv@gmail.com\\}
\and
\IEEEauthorblockN{Jordi Forn\'e, Celeste Campo, Ana Fern\'andez, Luis J. de la Cruz, Carlos Garc\'ia-Rubio, Andr\'es Mar\'in, Ahmad Mohamad Mezher, Daniel D\'iaz, H\'ector Cerezo, David Rebollo-Monedero, Patricia Arias, Francisco Rico}
}

\maketitle

\begin{abstract}
Major advances in information and communication technologies (ICTs) make citizens to be considered as sensors in motion. Carrying their mobile devices, moving in their connected vehicles or actively participating in social networks, citizens provide a wealth of information that, after properly processing, can support numerous applications for the benefit of the community. In the context of smart communities, the INRISCO  proposal intends for (i) the early detection of abnormal situations in cities (i.e., incidents), (ii) the analysis of whether, according to their impact, those incidents are really adverse for the community; and (iii) the automatic actuation by dissemination of appropriate information to citizens and authorities. Thus, INRISCO will identify and report on incidents in traffic (jam, accident) or public infrastructure (e.g., works, street cut), the occurrence of specific events that affect other citizens' life (e.g., demonstrations, concerts), or environmental problems (e.g., pollution, bad weather). It is of particular interest to this proposal the identification of incidents with a social and economic impact, which affects the quality of life of citizens.
\end{abstract}

 \begin{IEEEkeywords}
    Early detection of incidents, smart cities, citizen sensor, vehicular communications, big data analysis, social networks.
 \end{IEEEkeywords}

\section{Introduction}
\label{sec:intro}
\noindent

Nowadays citizens are seen as sensors in motion in the smart cities. They produce a huge amount of information from their smart-phones that will continue to grow exponentially when smart connected vehicles become a reality in the roads. The goal of our work is to quickly detect any unexpected situation in the city by properly analyzing this big amount of gathered data. The INRISCO (INcident monitoRing In Smart COmmunities) project \cite{INRISCO} does not intend to replace any  current emergency management service (e.g., 112 or 911), but rather to improve warning systems and complement them to manage unexpected situations which might have an impact in the citizens. In this regard, INRISCO aims to effectively address two key issues: (i) early detection of incidents and (ii) quickly dissemination of warning information. In the current technological context, it is feasible to obtain objective and subjective information from the new ``citizen sensor". Objective information is gathered through physical sensors set in their devices and in the smart city, whereas subjective information is gathered from opinions/comments posted on on-line social networks. The analysis of this massive amount of data (so-called Big Data) from structured and unstructured sources enables the early detection of unexpected situations as well as the estimation of their impact in the community. Going further, actuation is also possible by disseminating alert messages and suggested actions: globally, through on-line social networks; and locally, through ad-hoc or opportunistic networks dynamically built up in the community.

To ensure the citizens' participation, reliability, security, flexibility and privacy must be covered as requirements about which citizens are especially concerned. Firstly, it is crucial to ensure the accuracy of the information collected. Secondly, it is essential to ensure an adequate level of privacy to citizens, so that sensitive information (such as location, preferences, or other personal information) does not flow through social networks nor is collected by third parties without their explicit consent. Lastly, the proposal also cannot ignore the volume, velocity and variety of data provided by the citizens, as well as the critical nature of the detected situations and services involved.

For the integration of the main results of the project, INRISCO proposes a framework for the early detection of unexpected events and an efficient and effective dissemination of warnings. Our scheme has been deployed over a realistic scenario from a city, and over a dataset collected from a public on-line social network. 

The rest of the paper is organized as follows. An overview of the proposed system is presented in Section \ref{sec:scenario}. In Section \ref{sec:architecture} the INRISCO architecture is described. Section \ref{sec:DataCollection} provides details of the data collection process. In Section \ref{sec:FirstLayer}, we introduce the INRISCO layer in charge of the detection of any unexpected situation in the city. Section \ref{sec:Quality_Privacy} describes how INRISCO tackles issues of data credibility and data privacy. In Section \ref{sec:Sensing_Actuating} we show how INRISCO disseminates warning messages upon the detection of an incident. Related work is discussed in Section \ref{sec:related_work}. Finally, we conclude this article in Section \ref{sec:conclusions}, where we also we point out some future work.

\section{Description of the scenario}
\label{sec:scenario}
\begin{figure*}
	\centering
	\includegraphics[width=0.8\textwidth]{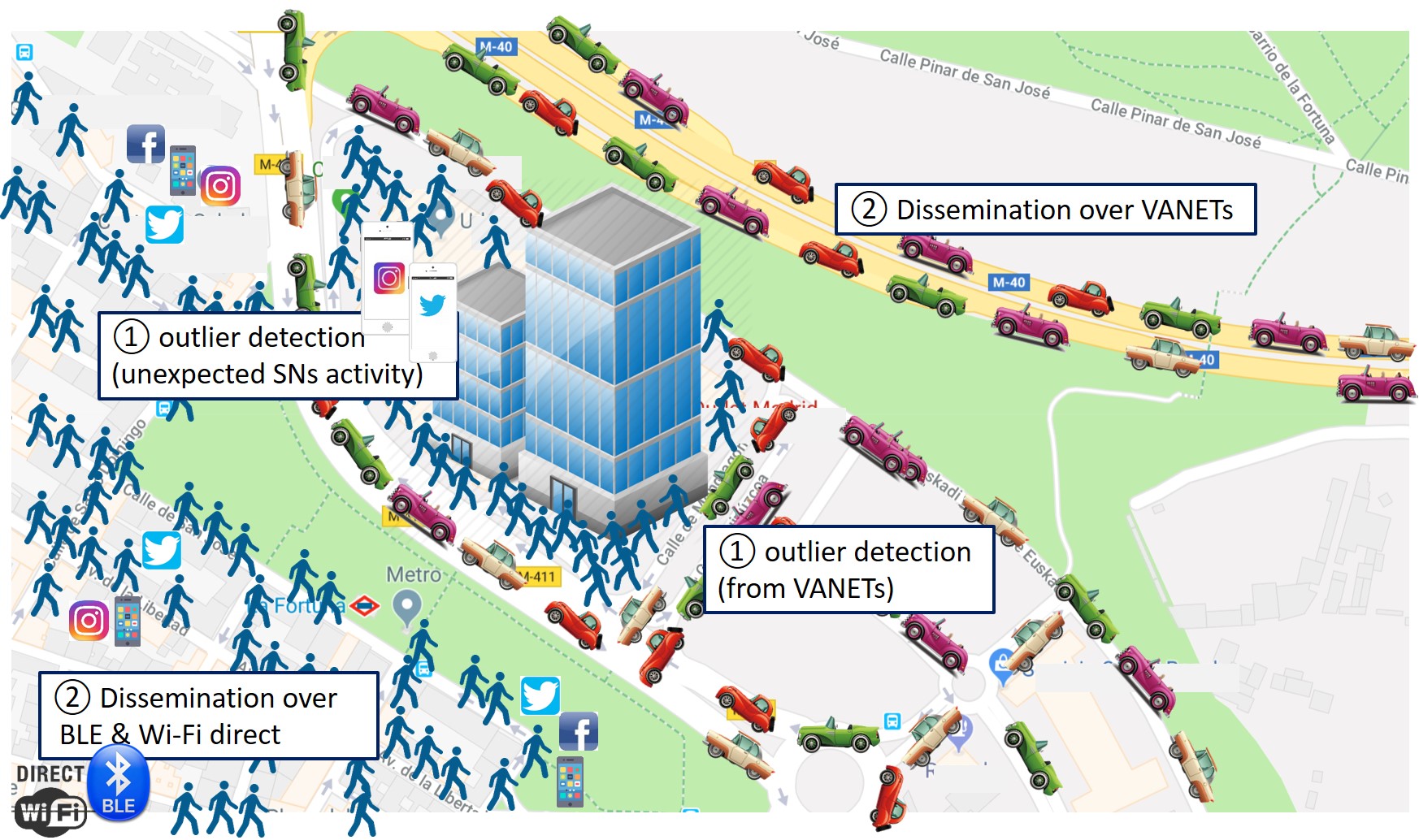}
	\caption{Scenario as case study of the INRISCO framework.}
    \label{fig:INRISCO_scenario}
\end{figure*}

\begin{figure*}
	\centering
	\includegraphics[width=0.85\textwidth]{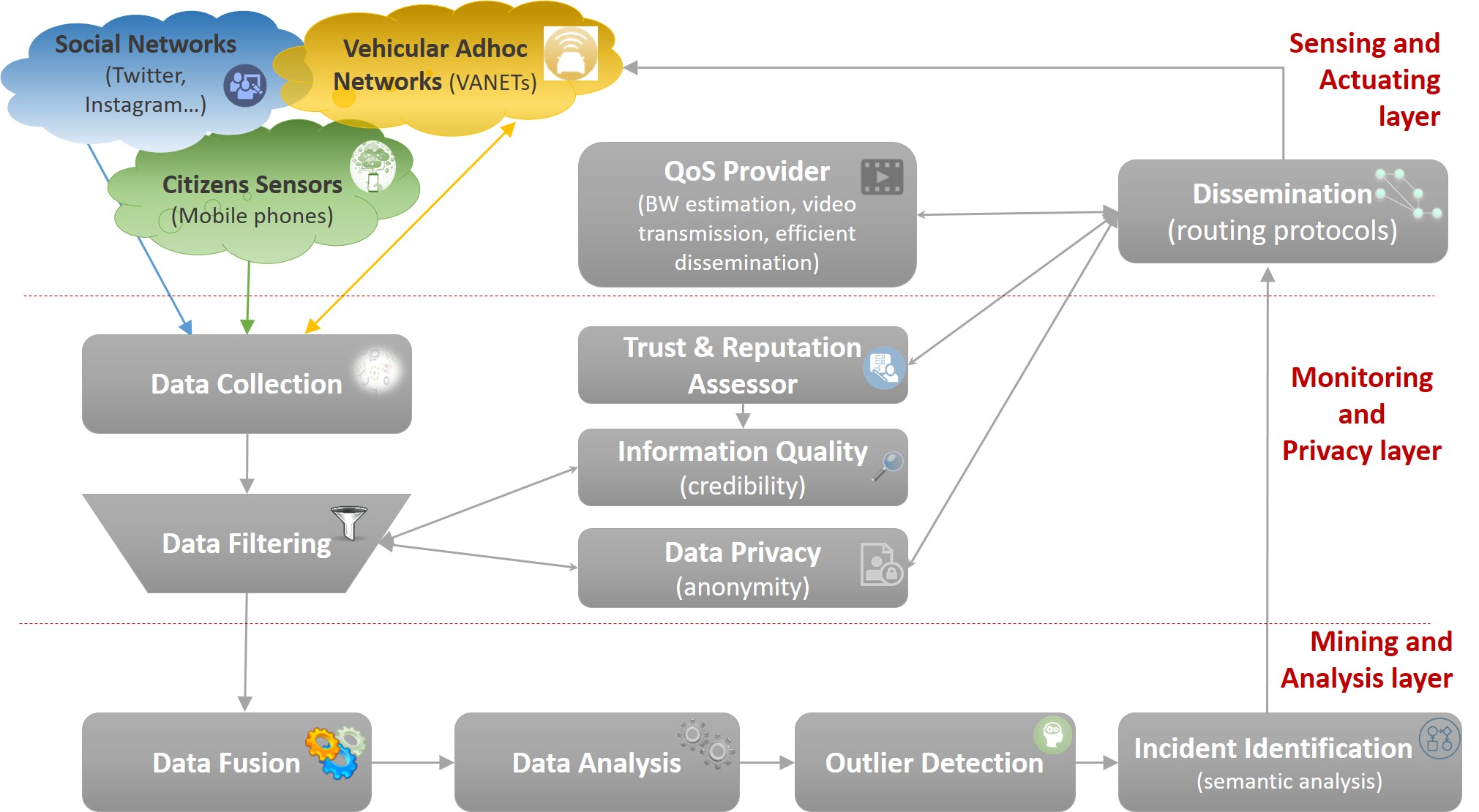}
	\caption{\label{fig:INRISCO_architecture}Architecture of the INRISCO framework.}
\end{figure*}

\noindent
The scenario to illustrate the INRISCO operation is depicted in Fig. \ref{fig:INRISCO_scenario} and its structure of layers is shown in Fig. \ref{fig:INRISCO_architecture}, where we can see the relationship between the different actors involved in the considered smart community.

In our proposal we consider that the city of Legan\'es (Madrid, Spain)  uses the INRISCO \cite{INRISCO} platform to detect any incident in the city based on different data sources like social media activity and exchange of messages in networks without infrastructure. More specifically, the data used in the INRISCO system is gathered from two public  location-based social networks (LBSN), Twitter and Instagram, and also from vehicular ad hoc networks (VANETs) and mobile ad hoc networks (MANETs). Regarding the former, the publicly shared posts in social media contain two kinds of data: (i) the GPS location of the posting device and (ii) the content of the post (natural text and/or images). Regarding the latter, we assume that the information obtained from VANETs and MANETs consists in multimedia warning messages about an incident (e.g., traffic jam, work zone in a street, blocked street, traffic accident, etc.) that might have an interest for other citizens (e.g., drivers, pedestrians). The message contains the GPS location and a text or video message regarding that incident.

Let us consider that today, after filtering and pseudonymizing the gathered posts, INRISCO has detected an unusual number of people near the Sambil shopping center of Leganés (see Fig.~\ref{fig:INRISCO_scenario}) and, consequently, an alert of a moderate crowd triggers \cite{Sambil}. This activation is a consequence of a quick analysis based on the entropy of Twitter and Instagram posts in the area that describes the pulse of the city, i.e. how citizens as a whole are moving all through Leganés. This global analysis triggers a second deeper and localized analysis: a density-based clustering system that identifies the locations of the crowds that have really changed according to the normal pattern of the city. After this second analysis, INRISCO confirms that a moderate and unexpected crowd is located in the Sambil area.

Automatically, a natural language processing module analyzes the post content and infers that something related to a {\em job fair} is located at the shopping area. With this information, INRISCO checks the foreseen events in the area to find that the job fair was already authorized by the council, so the alert is neutralized. The unexpected social media activity that feeds the INRISCO system is consequence of those citizens that are already in the job fair and who are posting pictures and texts describing the long queues to deliver their resumes to the different companies. Other citizens, trapped in their cars as consequence of the first traffic jams, are also posting the situation. Finally, companies also are publishing information about the same event in their social media profiles. 

On the other hand, and taking into account the unexpected impact of this event in the surroundings, the INRISCO system triggers the information dissemination procedure. For this purpose, the INRISCO system provides two different distribution mechanisms: firstly, a beaconing system that makes use of VANETs and MANETs to disseminate warning messages among citizens around, i.e., drivers and pedestrians, respectively; secondly, a beaconing system, based on Bluetooth low energy (BLE) communications, is used to spread interesting information for citizens nearby. INRISCO starts a dissemination procedure in the beacons near the shopping center area informing the citizens nearby about this event, just in case the job fair interferes with their normal life.

Thanks to this alert people like Peter, whose intention was going to the shopping area with his two children, changes his plan before entering the shopping center neighborhood. Alternatively, he decides to go to a different playground area. Peter, as thousands of citizens of Leganés, is connected to the beaconing system provided by the city and receives this alert on time. Another example is Eve, an amateur runner who normally trains in the park near the shopping center. That morning Eve receives the same alert than Peter. However, and since she is trying to find a better job, Eve decides to go to the job fair instead. There is also Alice, who usually drives near Sambil to go to her job. She receives the alert before reaching the traffic jam, so she decides to change her daily route and take the previous exit. 
Thanks to the detection and dissemination supported by INRISCO, the situation near the shopping center area is under control. Drivers not interested in the job fair took other different roads to get their destinations, while pedestrians not interested in the event decided to change their activity. As a consequence, the first traffic jams did not affect citizens not involved in the job fair. Additionally, people were informed about the event and those interested were able to participate.

\section{Background}
\label{sec:background}
\noindent

In this section we detail the technical background that establishes the bases on which we have developed our proposals. Mainly, we focus on pointing out the bases of (i) data collection and filtering, (ii) clustering techniques and (iii) data collection and dissemination using vehicular networks.

\subsection{Data Collection and Filtering}
\label{sect:backgDC}

Data collection is the process of gathering data or information, which can come from open sources (e.g., social networks (SN)) or sensors from mobile devices.

Social networks are a valuable data source, because these allow collecting a huge amount of information, through three main techniques for data acquisition: (i) network traffic sniffing, (ii) implementation of specific applications using an API for each SN, or (iii) crawling of the user social graph~\mbox{\cite{canali11}}. The second technique has become the most popular, because several social networks provide a set of APIs to acquire public data, e.g., Twitter, Facebook, Instagram, Forsquare, Google+, etc. Specifically, the Twitter API is a public REST API that provides access for reading and writing tweets or users’ data (i.e., location, followers, followed people, account creation date, etc.), using an OAuth access token for authentication. Responses are provided in JSON format. Twitter Search API limits the number of calls both per user and per application, whereas Twitter Streaming API allows accessing only the 1\% of the published tweets, and the kind of sampling is not public. Twitter also provides premium access (PowerTrack, Decahose, Firehose and Replay) based on the Streaming HTTP protocol to deliver data through an open, streaming API connection. Therefore, instead of delivering data in batches, consequence of repeated request by the client, a single connection is opened between the client and the API and new results are sent through that connection whenever new matches occur. Instagram, on another hand, used to have a public API that was shut down in 2018 and replaced by the so-called Graph API. Consequently, now third-party applications need to be approved by Instagram before they can access the Graph API, by using an Instagram Business Account to access the information.

In social networks, users freely communicate with each other and share ongoing activities, news, opinions, among others. So, it is possible to detect emerging events, analyzing such information. Likewise, location information allows analyzing individual's mobility patterns.

The individual's mobility patterns can be also gathered using cellular-based traces from users' mobile devices, because location information cannot be always obtained from data provided by social network API. For that, ad-hoc applications can be developed.

The collected data from social networks should be refined through a wide range of filtering strategies, e.g., removing redundant or impartial parts of data, evaluating owners' trustworthiness, scrub private fields, etc., in order to improve the data quality and reliability or preserve users’ privacy.

\subsection{Clustering techniques}
\label{sect:backgroundClustering}

Clustering is ``the process of partitioning a set of data objects into subsets or clusters such that objects in a cluster are similar to one another, yet dissimilar to objects in other clusters'' \cite{clustering}. Clustering methods are generally classified in four groups: partitioning approaches (where the number of clusters is pre-assigned), grid-based (where the object space is divided into a pre-assigned number of cells), hierarchical (where the data is organised in multiple levels) or density-based (where density notion is considered). We have applied two different strategies, a partitioning and a density-based algorithm, which will be used for different purposes in our approach. 

Density-based clustering algorithms organize the data in regions (clusters) where the elements are dense and which are separated by areas of low element density, which are considered as noise. These algorithms are able to (i) discover clusters of arbitrary shapes, (ii) handle sparse regions (which are considered as noise regions) and (iii) work without knowing the number of clusters in advance. Among the different proposals in the literature \cite{dbscancomparative}, we have selected DBSCAN ~\cite{dbscan} (density-based spatial clustering of applications with noise), since it is efficient and flexible while it does not add any extra-functionality neither any extra-computational load.

DBSCAN needs two parameters ($minPoints$ and $\epsilon$) to define the density of the clusters: the minimum number of elements ($minPoints$) that must be located within a given distance $\epsilon$ from an element in order to start forming a cluster. Tuning the two parameters of DBSCAN is essential for the algorithm to work properly, since the algorithm is very sensitive to both of them. In fact, DBSCAN defines an element as a noise point if it is not enough close to other elements, otherwise DBSCAN assigns the element to a particular cluster. For this, DBSCAN determines the local density at each element by using the previous two parameters: reachability parameter (distance value $\epsilon$) and the minimum number of elements ($minPoints$). An element or point $p$ which meets the minimum density criterion, i.e. there are at least $minPoints$ located within a distance $\epsilon$ from $p$ is considered a core point and defines an $\epsilon-neighbourhood$. Any element or point $q$ within the $\epsilon-neighbourhood$ is considered as directly reachable from $p$. Any element or point $q$ is reachable or connected by density from $p$ if there is a path of elements or points that connect both elements through chains of $\epsilon-neighbourhoods$. Therefore, a core point forms a cluster together with all elements or points that are reachable from it. As aforementioned, those points that are not assigned to any cluster are considered as noise.

\subsection{Data collection and dissemination using vehicular networks}
\label{sect:data_vehicular}

Once an incident has been positively identified, INRISCO disseminates proper emergency messages through social networks (e.g. Twitter, Instagram), MANETs and VANETs to warn citizens as soon as possible, see Fig.~\ref{fig:INRISCO_architecture}.

To design our proposals of routing protocols for VANETs, we used the well-known greedy perimeter stateless routing (GPSR)~\cite{GPSR} as a reference since it was one of the first geographic proposals over which many other proposals have been proposed. Vehicles are assumed to know their locations as well as the destination location. GPSR has two different modes to forward packets: (i) greedy mode, which is used by default; and (ii) perimeter mode used when it is not possible to use the greedy mode. This happens when there is no candidate node better than the current node that holds the packet. In greedy mode vehicles basically select the neighbor node closest to the packet's destination. In perimeter node, the current holding vehicle applies the right hand rule to find a possible next hop to forward the packet.

Several proposals have been presented in the literature to improve the basic GPSR. Specifically, in the literature we can find multimetric routing protocols that notably improve GPSR, which just uses the distance as the decision routing metric. For instance, our proposal multimedia multimetric map-aware routing protocol (3MRP)~\cite{TVT_Ahmad_2017} uses five metrics to optimize the selection of the best next forwarding vehicle (see section~\ref{sec:INRISCO_VANET_UPC}).

Data dissemination in VANETs is a well-studied topic. Among the main solutions that focus on urban scenarios is Urban Vehicular BroadCAST (UV-CAST) ~\cite{UV_CAST}, which uses digital map information to verify if the vehicle is at an intersection or not. This condition creates message broadcasts to other road directions. Additionally, UV-CAST can assign to more than one vehicle the responsibility for opportunistic forwarding, so vehicles can forward the message more than once. UV-CAST combines broadcast suppression and store-carry-forward, i.e., these proposals tackle the broadcast storm and the disconnected network problems simultaneously. According to the good results, this combination is an important basis for the development of a dissemination protocol for road safety applications.

To improve the data dissemination process some proposals derive the probability of forwarding a packet using game-theoretical algorithms. For instance, the work in ~\cite{Dissemination_1}  presents a technique to mitigate the broadcast storm problem through a game-theoretical mechanism based on “the volunteer's dilemma”. With this mechanism, the probability of a vehicle forwarding a message is computed considering the cost and the benefit of message propagation.

\section{INRISCO architecture}
\label{sec:architecture}
\noindent
The INRISCO architecture is composed by three layers, as Fig.~\ref{fig:INRISCO_architecture} shows: (i) {\em Sensing and Actuating}; (ii) {\em Monitoring and Privacy}; and (iii) {\em Mining and Analysis}. 

The {\em Sensing and Actuating} layer manages the information inputs (e.g., text, audio, video, location) from citizens that use their mobile devices to collect data from their environment (through physically integrated sensors), or from their interactions in social networks. In addition, data can be gathered from vehicular networks. This layer is also responsible for providing the appropriate technological basis to support dissemination of warnings or recommendation messages about detected incidents through VANETs and social networks. The dissemination will be carried out with emphasis in opportunistic communication and quality of service, by taking advantage of the users' mobility and location.

The inputs are collected and managed by the following layer, {\em Monitoring and Privacy}. In this second layer, data is collected in real-time to support the detection of evidence-based incidents. The collected data is filtered and anonymized in order to guarantee the data quality to be analyzed and the users' privacy, respectively. The filtering is based on trust and reputation information, which is obtained from the analysis of users' behavior through their traffic patterns and shared posts. The applied filters try to prevent gathering data from a same citizen with multiple identities, from untrusted users, hoaxes, {\em etc.} To enforce privacy, two main protection facilities are provided, namely sanitization and statistical disclosure control, as it is explained in section \ref{sec:INRISCO_PRIVACY_UPC}.

Finally, the {\em Mining and Analysis} layer is responsible for: (i) data fusion over the data streams extracted from the social contribution of citizens in social media, sensor data and/or public infrastructure; (ii) inferring the behavior of smart communities by using time-series data mining techniques; (iii) outlier detection for finding out differences from the expected patterns; and (iv) identification of the incidents by analyzing the content of the collected information. 

Each layer is described in detailed in the following sections according to the methodological order: data collection, data analysis and processing, quality \& privacy issues and, finally, dissemination aspects.


\section{INRISCO: Sources of data}
\label{sec:DataCollection}
\noindent

The {\em Monitoring and  Privacy} INRISCO layer depicted in Fig.~\ref{fig:INRISCO_architecture} is in charge of three essential tasks: (i) collecting data from citizens and vehicles, which are envisioned to act as real sensors of the city behavior; (ii) assessing the quality of the data (credibility) and (iii) protecting the users' privacy. In this section, we face the first aspect, whereas the other two are tackled in section \ref{sec:INRISCO_PRIVACY_UPC}. 

INRISCO mainly works with data obtained from three sources: social media, VANETs and cellular-based traces. 

On the one hand, social media sources usually provide a set of application programming interfaces (APIs) that can be used to gather data. Depending on the social source (Facebook, Instagram, Twitter, {\em etc}.) the pursued content may vary and the procedure needs from different permissions. For the subsequent modules to work properly, we have collected both geo-tagged posts and non geo-tagged posts. The former will be used  by the {\em Data Analysis} and {\em Outlier Detection} modules ({\em Mining and Analysis layer}) to detect unexpected behavior in specific areas along the urban areas. The latter will be used by {\em Incident Identification} module to supplement the geo-located analysis, in order to infer the causes behind the detected anomalies.

On the other hand, VANETs are used to gather information about the traffic state (i.e., sparse, fluid, dense, traffic jam) on the streets. Vehicles periodically interchange beacon messages with their neighbor vehicles within their transmission range. Also, we assume that vehicles periodically send their traffic reports to the closest road side unit (RSU) through the VANET using a proper multihop routing protocol. RSUs are city infrastructure used to communicate to the INRISCO platform. This way, INRISCO is aware of the vehicles' density in the streets of the city. Upon the detection of an incident in the street (e.g., an accident, a traffic jam) by the modules of this layer, the {\em Dissemination module} will act in consequence, and thus a warning message will be sent to those vehicles approaching the affected area. For instance, a crashed vehicle may disseminate warning messages through the VANET to alert close vehicles around the accident, aiming at avoiding another cascade accident. In this sense, the design of a smart dissemination protocol is crucial to avoid broadcast problems (see section \ref{sec:Sensing_Actuating_UPC}).

Finally, cellular-based traces give information about mobility at individual level, which unveils interesting information, since the subsequent locations a person visits define her in many ways. Continuously tracking the user and appropriately mining the resulting sequence of locations disclose a lot of information that can be used for early detection of abnormal situations in cities (i.e., incidents). We have obtained a dataset of these traces, which was collected in~\cite{phdalicia}, that contains mobility data from 25 users tracked by means of their mobile phones during more than one year. Among the participants, there are people living in five different countries and working/studying in completely different and independent places. Thus, they do not share the same space or specific timetable. This data was also collected using a baseline data collection scheme, i.e., collecting every cell change experienced by the device, and also including timestamps of incoming and outcoming calls.

\section{Mining and Analysis Layer}
\label{sec:FirstLayer}
\noindent
The {\em Mining and Analysis} layer is the core of the INRISCO system, since it supports the detection of unexpected events in urban areas and feeds the {\em Dissemination module} with information to be distributed mainly in the surroundings of the aforementioned events, as Fig.~\ref{fig:INRISCO_architecture} shows. 

The first module of this layer, {\em Data Fusion} obtains information from different sources, mainly social media, cellular-based traces and VANETs, as explained in the previous section. The second and third modules of this layer, {\em Data Analysis} and {\em Outlier detection} focus on processing this data with the aim of detecting anomalies. Finally, the fourth module {\em Incident Identification} tries to know, by applying natural language techniques, the underlying reasons that could justify the aforementioned anomalies.

In this section, we firstly summarize (section \ref{sec:DBSCAN_UVIGO}) the methodology we proposed to identify behavioural patterns in urban areas and how to apply these patterns to detect unexpected behaviors all around the city, as a whole, without focusing on specific users or citizens. Then, we detail some studies that we have performed to detect anomalies but using individual movements (section \ref{sec:individual_movement}). Finally, we describe the natural language techniques used to know the underlying reasons that could justify those identified anomalies (section \ref{sec:DBSCAN_UVIGO}).

\subsection{Identifying behavioral anomalies: a DBSCAN-based methodology}
\label{sec:DBSCAN_UVIGO}

\noindent

With the aim of analyzing the city or a specific urban area as a whole to detect unexpected changes in the activity, i.e. in the city pulse, we take advantage of the public geo-tagged posts that citizens share in social media. Our methodology combines social data mining, density-based clustering and outlier detection into a solution that can operate on-the-fly. The approach is based on gathering and analyzing information of prior locations of groups of people at different time slots all over the city. After analyzing this data, we have a reference pattern and, consequently, it is possible to check and compare real time data to detect abnormal behaviors in the city. Since identifying crowds is the main purpose of this work, the methodology demands a clustering algorithm, in which clusters are formed with points in dense areas whereas points in sparse areas are discarded. The clustering algorithm which best fits our purposes is density-based spatial clustering of applications with noise (DBSCAN) \cite{dbscan}. DBSCAN can find an unspecified number of clusters with arbitrary shapes, and also remove noise points, just setting two parameters ($minPoints$ and $\epsilon$) to define the density of the clusters: the minimum number of elements ($minPoints$) that must be located within a given distance $\epsilon$ from an element in order to start forming a cluster ~\cite{dbscancomparative}.

\begin{figure*}
	\centering
	\includegraphics[width=0.75\textwidth]{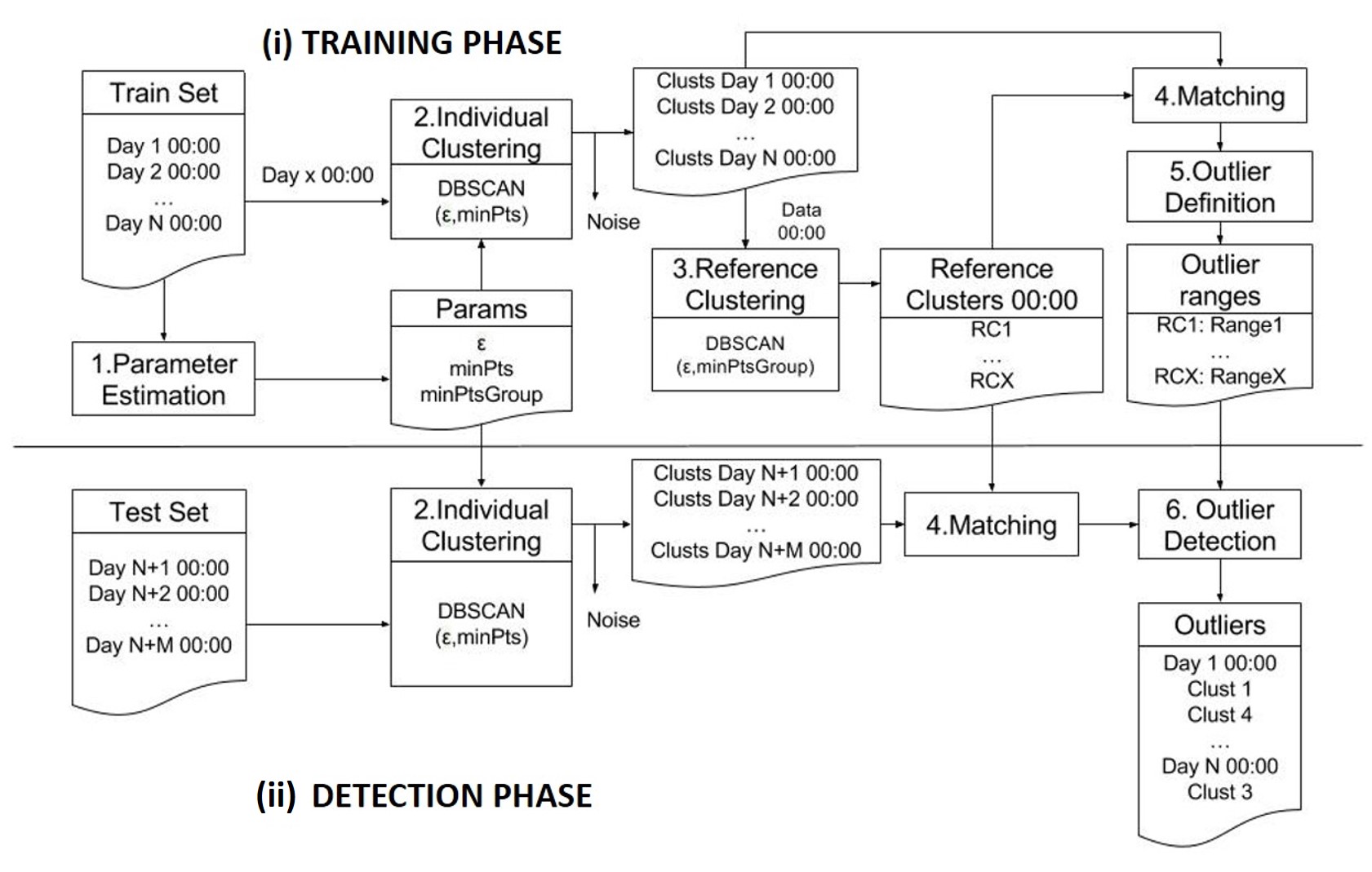}
	\caption{\label{fig:UVIGO:Methodology} Methodology to identify anomalies in the citizens' activity.}
\end{figure*}

Taking DBSCAN as our clustering algorithm, the complete methodology consists in two main phases: (i) {\em Training} and (ii) {\em Detection}. On the one hand, the training phase tries to obtain the location of the reference clusters (RCs) for an average day and also it tries to define the expected size of the clusters which match each of these RCs. In other words, it obtains the models which represent the general behavior of the citizens. This phase requires combining data from several days, and it is only performed once, before starting the detection phase. On the other hand, the objective of the detection phase is obtaining the clusters for each individual day and decide which of those clusters have an unexpected number of points in comparison to the number of points expected for the RC they match. This phase is performed on-the-fly as soon as the data is available. For this analysis, days are divided in intervals (15 or 30 minutes, for instance), and the results are calculated for each interval independently. This way, as soon as the data from a new interval is extracted the detection phase can be performed with this data. Also note that, since the RCs obtained in the training phase tries to represent the average behavior of all days in the data set, those days need to be similar (for instance, same day of the week, or any other suitable aggregation rules).  Therefore, the location of the RCs should represent the location of the usual activity in the urban area. For instance, RCs will be located (i) nearby business centers from Monday to Friday during typical working schedule, (ii) around sport centers and leisure areas after work, and (iii) nearby restaurant and bar areas at evenings and specially at weekends. In the use case of this paper, the shopping center area would have a cluster in the RC, but with no so much activity that reflects the usual social media activity any morning in the Sambil shopping center (no so much populated). Consequently, the alert is triggered because the system detects an unusually high activity in social media, compared to the usual one defined in the RC. This is because of the job fair that is taking place in the shopping center \cite{Sambil}. Fig.~\ref{fig:UVIGO:Methodology} overviews these two phases which correspond with the modules {\em Data fusion}, {\em Data Analysis} and {\em Outlier Detection} depicted in Fig.~\ref{fig:INRISCO_architecture}.

The details of the training and detection phases are explained in \cite{DOMINGUEZ2017} \cite{coin2019}, and summarized as follows. For the {\em Training phase}, (i) in Fig.~\ref{fig:UVIGO:Methodology}, the city model is obtained of-the-record previous to the detection phase, analysing the data from multiple days in the training set, by applying the following procedure for each one of the intervals in which we divided all the similar days:

The specific methodology used in INRISCO to tune the value of the two parameters ($minPoints$ and $\epsilon$) was detailed in [12], where we show that the algorithm is very sensitive to the appropriate selection of both parameters.

\begin{itemize}
	\item Parameter estimation: Obtain the necessary parameters to apply DBSCAN to each day in the dataset and obtain the parameters to apply DBSCAN to the average day of all similar days (same day of the week in our approach).  The specific methodology used in INRISCO to tune the value of the two parameters ($minPoints$ and $\epsilon$) was detailed in \cite{DOMINGUEZ2017}. The selection of the parameter is key, since the DBSCAN algorithm is very sensitive to them.
	\item Individual clustering: Use DBSCAN to obtain the clusters for each individual day and discard noise points.
	\item Reference clustering: Use DBSCAN again with data gathered from all days together. Thus, parameters now adapt to the fact that there are more days, and so more data points.
	\item Matching: Match individual clusters with RCs based on the distance between them. This distance, whose details are explained in \cite{DOMINGUEZ2017} measures the similarity level between two clusters by comparing the location of their centroids and also the density and extension of both clusters.
	\item Outlier definition: Define the outlier limits for each RC based on the number of points that all the individual clusters have.
\end{itemize}

For the {\em Detection phase}, (ii) in Fig.~\ref{fig:UVIGO:Methodology}, we use the parameters {\em Reference Clusters} and {\em Outlier Limits} obtained during the training phase, as well as the data from the Test set, which is obtained and analysed on-the-fly. This means that as soon as the data from a time interval is collected in real time, it can be instantly analysed to detect the outliers by performing the following steps:

\begin{itemize}
	\item Individual clustering: Use DBSCAN using the parameters estimated from the Training set.
	\item Matching: Match individual clusters from the Test set with RCs obtained from the training phase. 
	\item Outlier detection: Compare the number of points in each individual cluster with the limits defined for the RCs they match. 
\end{itemize}
	
This methodology was applied in different contexts, like in New York City (NYC) \cite{DOMINGUEZ2017}, where a dataset was obtained gathering data from Instagram during 22 weeks, obtaining the number of geo-located posts detailed in Table  \ref{table:weekPosts}.

\begin{table}[!htbp]
\centering
\caption{Post distribution per day}
\label{table:weekPosts}
\begin{tabular}{lrr}
  \hline
 Day & Total & Daily avg\\ 
  \hline
Mon & 506,479 & 23,021.77 \\ 
Tue & 495,620 & 22,528.18 \\ 
Wed & 489,545 & 22,252.05 \\ 
Thu & 490,591 & 22,299.59 \\ 
Fri & 508,247 & 23,102.14 \\ 
Sat & 503,879 & 22,903.59 \\ 
Sun & 503,851 & 22,902.32 \\ 
\hline
 & 3,498,212 & 22,715.66\\

   \hline
\end{tabular}
\end{table}

\begin{figure}[!htbp]
 \centering
    \subfigure[Reference cluster at 17:30\label{fig:pattern}]{\includegraphics[width=0.35\textwidth]{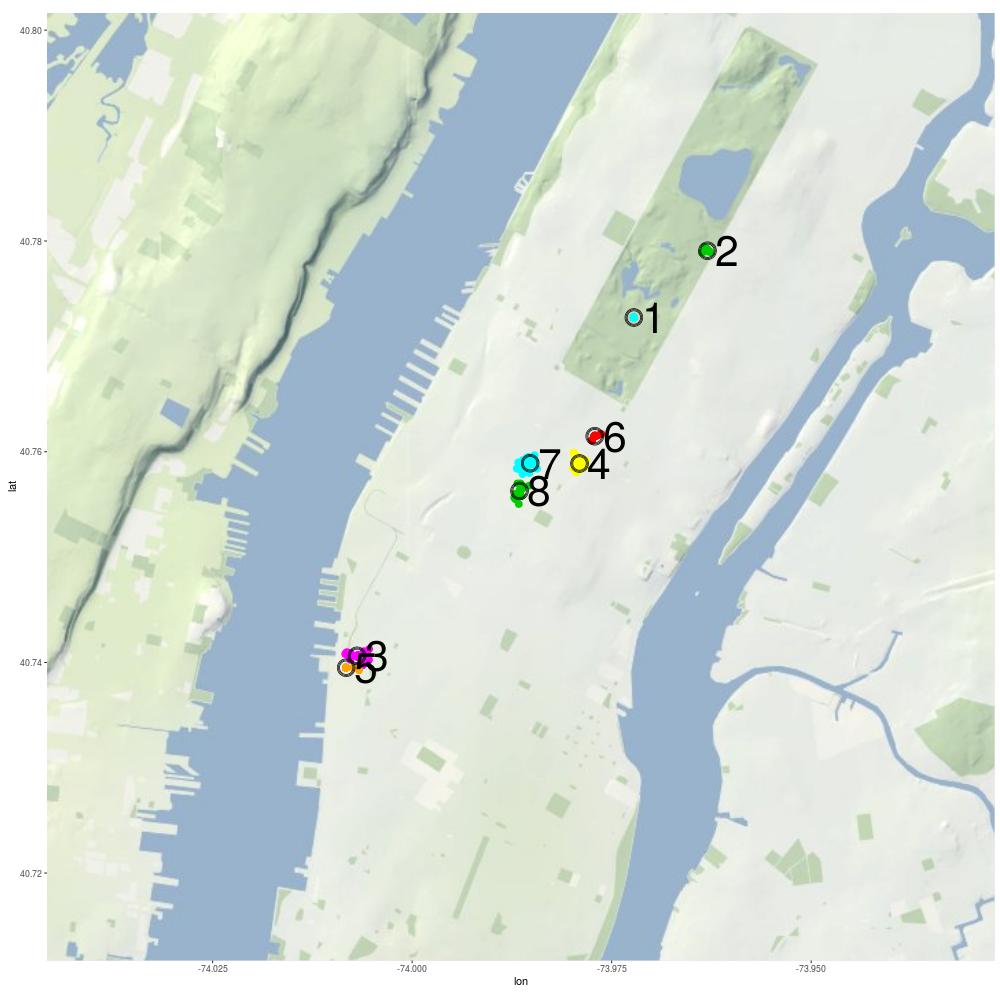}}
    \subfigure[Comic Con Day at 17:30 \label{fig:comiccon}]{\includegraphics[width=0.35\textwidth]{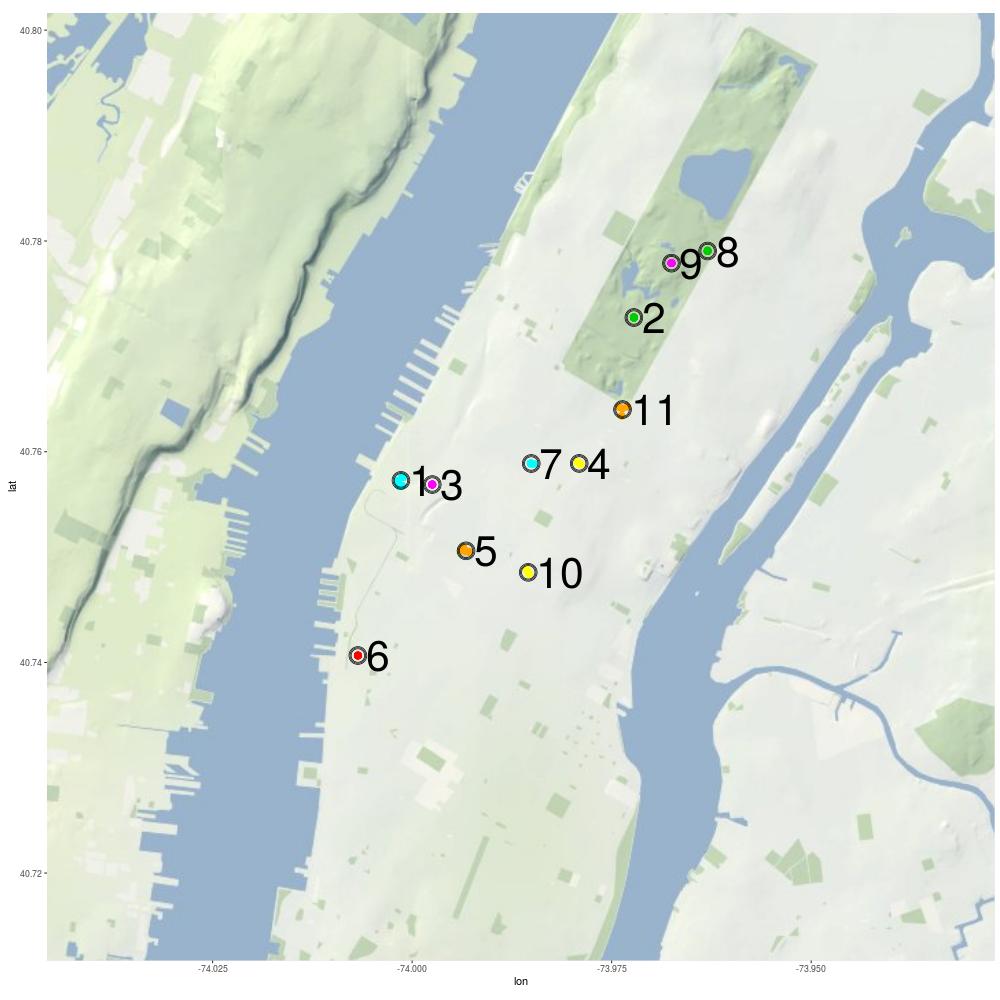}}
    \subfigure[Jonas Storm Day at 17:30\label{fig:jonas}]{\includegraphics[width=0.35\textwidth]{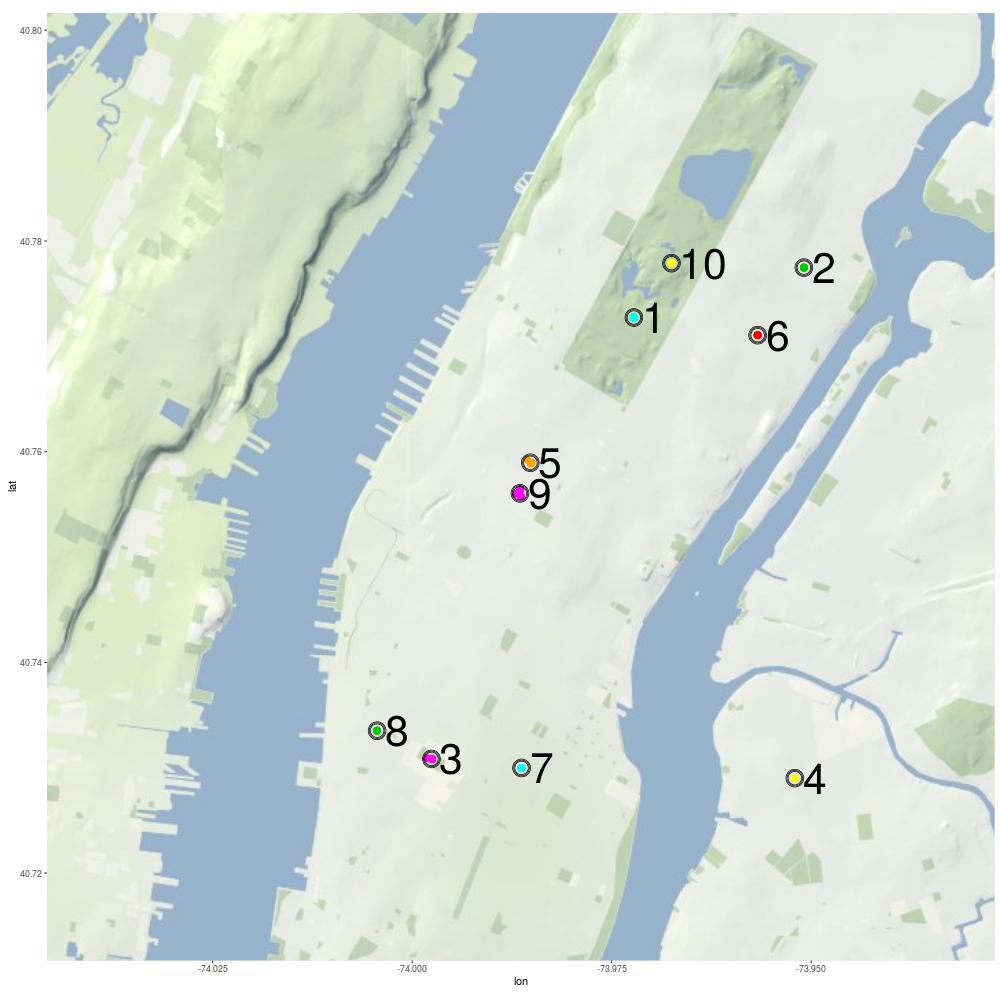}}
 \caption{Individual clustering for two special days vs. the reference cluster (expected behavior). Manhattan area, NYC.}
 \label{fig:individual}
 \end{figure}

After performing the training part of the methodology, we obtained a set of different patterns of the area under study showing the activity in social media, i.e. the pulse of the city: each pattern represents the city behavior each 30 minutes, and the patterns are obtained aggregating the data each day of the week. So, we defined 48 patterns per day of the week and 336 patterns in total. Figure \ref{fig:pattern} shows the reference clustering (pattern) typical of a Saturday at 17:30. Notice that the dots' colors are just used to distinguish clusters, with no further special meaning. After performing this training phase, the system is ready to monitor the daily activity using the activity in social media, Instagram in this case, detecting the differences between the expected behaviour (pattern) and the current one. The results obtained allow us to identify anomalies, like the clustering in Figure \ref{fig:comiccon} that shows a different clusters pattern with higher densities than expected in specific areas in the city. Another example is Figure \ref{fig:jonas} that also shows different locations of the clusters and lower densities than expected. After checking the former, we realized that the Comic Con event (annual NYC fan convention dedicated to comics) was at the detected area. In the same way, after checking the latter we realized that this was the result during the Jonas Storm (historic blizzard in January 2016 in United States). 

\subsection{Analyzing the impact of individual movements}
\label{sec:individual_movement}
\noindent

We studied the impact of cellular-based location collection schemes on observed human mobility features~\cite{PERCOM2014}. We identified the most basic mobility features: (i) amount of movement (number of cell changes made per day), (ii) variety of visited locations (how many different locations are visited by the user per day), (iii) visits distribution (fraction of visits concentrated by each location), (iv) randomness (uncertainty we have about the next event in a sequence), and (v) predictability (bound of the maximum percentage of correct predictions the best algorithm could ever attain). We observed that the traces based on data stored in the network nodes themselves, recording the base transceiver station (BTS) to which the device is connected when the user is making or receiving a call, or sending or receiving a short message, known as call-detailed records (CDR), lead to biased mobility indicators~\mbox{\cite{PERCOM2014}}.

In ~\cite{PERCOM2015} we proposed that randomness in people's movements might serve to detect behavior anomalies (i.e., a weird behavior). The concept of entropy can be used for this purpose, but its estimation is computational intensive, particularly when processing long movement histories. Moreover, disclosing such histories to third parties may violate user's privacy.
With the goal to keep the mobility data in the mobile device itself yet being able to measure randomness, we proposed a fast entropy estimator scheme, based on Lempel-Ziv (LZ) prediction algorithms, intended to work sequentially with a low computational cost. This algorithm allows us to rapidly detect changes in the normal mobility routines of a user, and can be used for tracking users or processes where mobility pattern changes need to raise a warning sign.

However, tracking user mobility rises privacy concerns. In ~\cite{ENTROPY2015} we analyzed privacy-enhancing mechanisms based on information theory concepts, such as entropy, applied to locations and mobility profiling scenarios.  We have shown that the theory applicable to these low alphabet cardinality and memoryless processes, cannot be directly applied to more complex cases such as the mobility profiles of users.  In complex (real life) cases, little privacy enhancement is observed, unless utility is completely lost. In this sense, the fast entropy estimator scheme we proposed in ~\cite{PERCOM2015} can be executed in the mobile devices instead of sending data to third parties, so preserving the privacy of the users. We explain more details about the INRISCO treatment of users' privacy in section (\ref{sec:INRISCO_PRIVACY_UPC}).

\subsection{Incident Identification}
\label{sec:GRADIANT}
\noindent

Once a crowd has been detected, the {\em Incident Identification} module (see Fig.~\ref{fig:INRISCO_architecture}), tries to find out the reasons behind the detected crowds (e.g., celebrations, sports events, festivals, demonstrations) by analyzing the textual content of the collected social media data. In the previous phases, the analysis only pay attention to the location (latitude, longitude) of the collected posts. However, the content of those posts also offers a really valuable information: the comments and opinions shared by the users. Assuming this information is usually related to the physical location from where the post is shared, its analysis enriches the knowledge about what is happening at that point and at that moment.

With this aim, we have performed different approaches to know the underlying reason that may justify any anomalous behavior detected in the urban area. Firstly, in \cite{GRADIANT-UVIGO}, we performed a light analysis of the topics based on a bag-of-words model corresponding to the fictitious document obtained by merging  all posts in the region under study and during the period under study (global topic) or all posts grouped in an identified crowd (localized topic). In both, the model represents the topic as a bag of posts' words, disregarding grammar and even word order although keeping multiplicity i.e., maintaining the number of times each word appears in the data set. Before the bag-of-words model, posts are pre-processed by lowercasing and by removing punctuation, numbers and stop-words. Then, the document term matrix (DTM) and the term document matrix (TDM) can be constructed to capture the frequency of terms and to apply several kind of analysis, for example, clustering, classification and association analysis. As mentioned, we report topic analysis at two level. (i) At the level of the global topic of the event, we focus on the document constructed from the posts all over the region and all over the day. Then, unsupervised analysis is applied by clustering the words. (ii) At the level of the localized topic of the event, the focus is the tag cloud of each detected crowd so we can generate a tag map of the topic discovered in each crowded location.

However, we also faced other solutions that require a deeper analysis of the text in the posts gathered from social media: (i) the \emph{early story detection} module, which facilitates identifying the cause of an incident occurring in a certain area of the city  \cite{GRADIANT-UVIGO}, and (ii) the \emph{incident classifier}, which provides a supplementary approach for early incident detection. These analysis empowers INRISCO with filtering and explanatory capabilities of the unexpected events highlighted by the anomaly detection algorithms. Both models are to some extent language independent. The \emph{early story detection} is ready to operate in any language. The \emph{incident classifier} imposes more language restrictions because it needs training data that should be gathered and annotated by humans, who are needed to generate the models.

\subsubsection{Early story detection module}
\noindent
When INRISCO detects a potential incident in a certain location, by identifying changes in the movement of the crowds according to the normal pattern of the city, it triggers the \emph{Early story detection} module (see \emph{Incident Identification} in Fig.~\ref{fig:INRISCO_architecture}).

The \emph{early story detection} (ESD) module analyzes the text in the posts shared in social media during the period of the anomaly, by dynamically grouping citizens' posts in threads (the so-called \textit{stories}) according to their semantic similarity. To this aim, it first translates the post content to a mathematical representation in $\mathbb{R}^n$, i.e. the set of all vectors with $n$ real components. This vector representation is built in such a way that semantic-related content is clustered together. Hence standard similarity metrics between vectors could be employed to obtain the relatedness score (i.e., a numerical value) used to quantify the degree to which two concepts are associated with each other. Here we summarize the process used in INRISCO to detect anomalies and we point out references for further details.  We have employed a character-based vectorization \cite{broder1997resemblance}, although corpus based vectorizations are also possible within this architecture \cite{gabrilovich2007computing,cilibrasi2007google,turney2001mining}. Under this setting, standard similarity metrics to compare vectors could be employed to obtain the relatedness score used to quantify the degree to which two concepts are associated with each other. In this case we have employed cosine similarity that measures the angle between two vectors but other common valid options are the $L_1$ (Manhattan) and $L_2$ (Euclidean) distances. Considering two vectors $\vec{a}$ and $\vec{b}$, the well-known cosine similarity is given by the following expression:

\begin{equation} \label{eq:cosine_similarity}
cos(\vec{a},\vec{b}) = \frac{\vec{a} \cdot \vec{b}}{\|\vec{a}\| \cdot \|\vec{b}\|}
\end{equation}

Higher values means the concepts are similar (results are between $-1$ and $1$). Calculating all the similarity metrics between pairs of points is very time consuming ($\mathcal{O}(n^2)$ complexity). To be able to process content data at a high rate, the model selectively picks the best candidates from historic data, limiting the number of comparisons to perform to just a few. We have employed multiple locality sensitive hashing (LSH) instances \cite{indyk1998approximate,gionis1999similarity} for this purpose that were able to process data in streaming, returning only the candidates that are frequently clustered together. The distance metric is only calculated between one point and a small subset of posts boosting the analytical speed without noticeable degradation in performance (between $50$ and $100$ are typically enough). Finally, for each post content, the module decides between creating a new story or adding the new post to an existing one using a similarity threshold. Although the relatedness score is the most influential in this decision, the module could consider additional ad-hoc criteria (e.g., active time of the story or geolocation data).

From the outcome, it is possible to explore the most relevant stories (i.e., those topics having the greatest impact in the city during the potential incident), the relations among the users involved in them, as well as the specific ones posted from the location the anomaly has been identified. The latter are specially significant when the anomaly corresponds to a relevant decrease in the number of people around the area.

This module demands no supervision. Thus, it could be deployed easily in different scenarios. It starts filtering content once it is plugged to a social data stream and could be executed both over continuous flows or historical corpus of data. Thanks to its filtering capabilities, the results gathered with it could be either plugged to interactive dashboards for analysis of historic data or to real-time automatic systems which are more computationally demanding and would not be able to process all the raw generated content otherwise. 

\begin{figure}[htb]
	\centering
	\includegraphics[width=.8\linewidth]{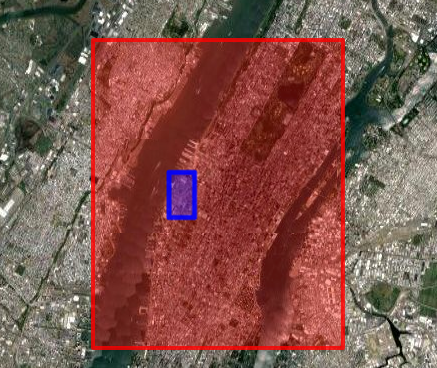}
  \caption{\label{fig:comicconarea} Comic Con area within the NYC geolocated dataset.}
\end{figure}

The ESD has been evaluated in different contexts, such as the NYC dataset presented in Section \ref{sec:DBSCAN_UVIGO}. In order to validate the detection of the Comic Con event, we restricted the area under analysis as Figure~\ref{fig:comicconarea} shows. Also, we limited the time to the weekend where the event took place. We performed the experiment changing the similarity threshold value ($0.6$, $0.65$, $0.7$ and $0.75$). The results, as  Table~\ref{threads:comic:in} shows, were consistent. The ESD extracted meaningful stories with information about the Comic Con within the conference's sub-area of influence, regardless of the selected threshold.

\begin{table*}[htb]
\begin{center}
\caption{\label{threads:comic:in} Top-three threads under the area of influence of the NYC Comic Con for different threshold values}
\begin{tabular}{|c|l|c|}
\hline \bf Threshold & \bf Story Threads & \bf Size \\  \hline

\multirow{3}{*}{0.6} & \parbox[t]{8cm}{\textit{Found Xur before the weekend ended along with Eris. \#destiny \#xur \#nycc \#nycc2015 \#newyorkcomiccon}} & 324 \\ \cline{2-3}
 & \parbox[t]{8cm}{\textit{\#Gaming \#NYCC2015 \#COMICCON}} &  320\\ \cline{2-3}
 & \parbox[t]{8cm}{\textit{Chewbacca \#starwars \#NYCC \#NYCC2015}} & 194\\
 \hline 
 
\multirow{3}{*}{0.65} & \parbox[t]{8cm}{\textit{\#Gaming \#NYCC2015 \#COMICCON}} & 434 \\ \cline{2-3}
 & \parbox[t]{8cm}{\textit{Chewbacca \#starwars \#NYCC \#NYCC2015}} &  162\\ \cline{2-3}
 & \parbox[t]{8cm}{\textit{Comic 411 at Day 2 of \#nycc@newyorkcomiccon \#fan220 @fan220dotcom \#comic411Photo taken by @sonnysofrito}} & 41\\
 \hline 
  
 \multirow{3}{*}{0.7} & \parbox[t]{8cm}{\textit{There goes another Clark lol  \#ComicCon ...}} & 179 \\ \cline{2-3}
 & \parbox[t]{8cm}{\textit{\#Gaming \#NYCC2015 \#COMICCON}} &  154\\ \cline{2-3}
 & \parbox[t]{8cm}{\textit{Chewbacca \#starwars \#NYCC \#NYCC2015}} & 145\\
 \hline 
 
 \multirow{3}{*}{0.75} & \parbox[t]{8cm}{\textit{\#Gaming \#NYCC2015 \#COMICCON}} & 440 \\ \cline{2-3}
 & \parbox[t]{8cm}{\textit{Chewbacca \#starwars \#NYCC \#NYCC2015}} &  167\\ \cline{2-3}
 & \parbox[t]{8cm}{\textit{Comic 411 at Day 2 of \#nycc@newyorkcomiccon \#fan220 @fan220dotcom \#comic411Photo taken by @sonnysofrito}} & 41\\
 \hline 
 
\hline
\end{tabular}
\end{center}
\end{table*}

\subsubsection{Incident classifier}
\noindent
Every text gathered from social media is analyzed in real time by the INRISCO's Incident classifier, within the \emph{Data Analysis} block (see Fig.~\ref{fig:INRISCO_architecture}). The output of the classifier is a tag indicative of the category in which the post is included. This module works at message level. That means, every single post is processed regardless previous content. The system receives the post content as input, preprocesses it to extract meaningful features and finally applies the classifier to extract the tag and the confidence. The confidence is a measure of the reliability of the model according to its own decision and it could be employed to filter those decisions with poorer confidence.

This is a supervised system that requires manually generated training data. The module has been trained to automatically classify a post in one of the following categories: 

\begin{itemize}
	\item \texttt{inclement\_weather}, including inclement weather conditions and related issues (e.g., floods).
	\item \texttt{traffic\_incident}, to identify texts containing mentions to jams, accidents and other traffic incidents.
	\item \texttt{crowds}, to detect references to crowds and possible causes (such as demonstrations, concerts or sporting events)
	\item \texttt{criminal\_incident}, texts which make reference to some criminal activity, vandalism, {\em etc}.
	\item \texttt{safety\_incident}, events such as accidents that may potentially influence city dynamics or concert citizens safety or structural changes in the city (e.g., a natural-gas explosion or a house fire). 
	\item \texttt{none}: none of the above. Posts that should be filtered out.
\end{itemize}

The classifier is a deep learning multilayered engine with multiple configuration modes which adapts to several input/output formats. For example, if the preprocessing steps include only tokenising the post content (i.e., splitting data into words) the system would employ a different internal configuration than when the system employs more complex inputs (e.g., syntactic/dependency analysed content). This configuration is transparent to the users of INRISCO ecosystem.

The \emph{Incident classifier} assists in the detection of anomalies and it is used both standalone and jointly with other anomaly detection systems in INRISCO. On the one hand, the module analyzes in real time every post gathered from social media, providing a supplementary approach for early incident detection. On the other hand, the classifier analyzes the stories inferred from the data related to a potential incident in a certain location -detected from changes in the movement patterns of the citizens (see section  \ref{sec:DBSCAN_UVIGO}). The tags associated with each post can be used to automatically filter uninteresting anomalies which are not topic-relevant for INRISCO (i.e., those events not directly related with city dynamics such as celebrity or company information which might be relevant for a city but does not involve people circulation).

\section{Monitoring and Privacy Layer}
\label{sec:Quality_Privacy}

This layer faces two very important tasks: (i) analyzing the quality of the information already gathered from the different sources of INRISCO and (ii) assuring privacy to the users' data.

\subsection{Information Quality}
\label{sec:quality}

\noindent
This module is responsible for filtering the collected data in order to avoid of misleading information. It addresses the credibility in the INRISCO project, for preventing the dissemination of false rumors or fake news, which are gaining more significance nowadays. This module also manages underlying information security.
 
Several researchers have identified the powerful and collectively intelligence force of citizens as digital volunteers, but at the same time, it is well-known that the lack of credibility from social networks reduces their use for situations with a high sensitivity degree as are public incidents ~\cite{Gupta12}. For these reasons, we have assessed the credibility of the gathered information.  Credibility has been mainly tackled from two perspectives: (i) analyzing the trustworthiness of shared information by using Tweets and their attributes; (ii) verifying the source authenticity by using users' profiles in social networks and trust and reputation systems in VANETs.

For analyzing the trustworthiness of shared information, we have focused on Tweets' content and their metadata (i.e., author verification, account type, followers, friends, age, location). The Tweets' content is analyzed semantically to determine its relation with the context (i.e., the incident). Besides, from the tweet's content we obtain attributes as length, word numbers, punctuation marks, URLs number and hash tags. The account type allows us to establish a users segmentation: common, corporate or organization. The information related to followers and friends allows us to calculate ratios Tweets/followers and friends/followers. Using all of these information, Tweets are firstly classified in four categories -by an expert-: \textit{related to the context and with info} (R1), \textit{related to the context, but without info} (R2), \textit{non-related to the context} (R3) and \textit{skip} (R4), respectively. This classification is used to build a graph, which is used as an initial decision tree. Then, using Tweets in the first category (R1), other learning algorithms (e.g., K-nearest neighbor (KNN), Random forest, Multilayer Perceptron (MLP)) are applied to improve the results, by carrying out a second classification, according to credibility levels: credible, less-credible, non-credible. 

\begin{figure}[htb]
	\centering
	\includegraphics[width=\linewidth]{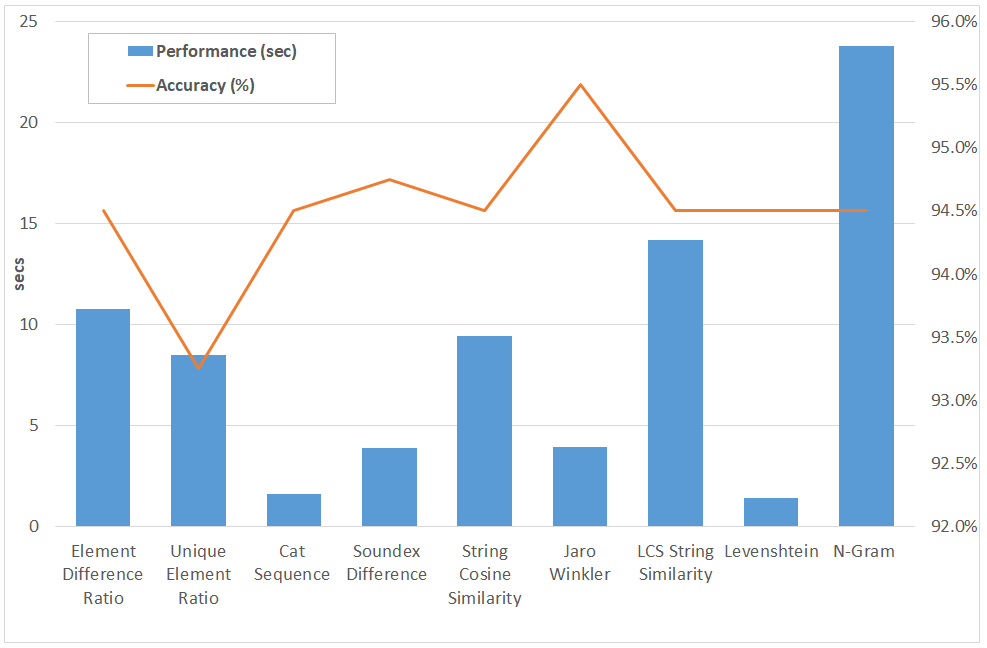}
  \caption{Performance and accuracy comparison on string matching algorithms.}
  \label{fig:profile}
\end{figure}

For source authenticity, we propose complementary lines related to the search of fake profiles and risk management. The search of fake profiles is based on collecting and comparing profile information (e.g., usernames, real names, locations, languages, verification, {\em etc}.) from different social networks.  In~\cite{Alvarez17}, we developed various string similarity algorithms for ``profile matching'': \textit{Element difference ratio}, \textit{Unique element ratio} and \textit{Cat sequence}. These algorithms were compared with other algorithms -in the literature- that require training. The algorithms proposed allowed us detecting duplicated identities or fake identities with a good accuracy. When the same user is found in different social networks, the system has a better user's knowledge. Using profiles from famous people, our preliminary conclusions showed certain feasibility and improvements of performance (in seconds) with similar accuracy (about 95\%), as shown in Fig.~\ref{fig:profile}. In the graph, Cat sequence showed results similar to Levenshtein, being the algorithms with the best performance. The algorithm that got best accuracy was Jaro Winkler (1\% more), but its performance was more than twice as good, compared with Cat sequence and Levenshtein. It is afford to mention that these results do not consider the training time.

Regarding risk management, we studied validation credentials for users' authentication in changing environments ~\cite{Alm16, Hin186}. The trust management is included to enhance authentication decisions. The approach involves new metrics to assess the security risks. This proposal allows INRISCO supporting security mechanisms that will be required for the interactions.


\subsection{Data Privacy}
\label{sec:INRISCO_PRIVACY_UPC}
\noindent
The collection of a wealth of data from social media and a rich variety of sensors in modern city infrastructures, prior to its analysis by automated information systems of increasing sophistication, is undoubtedly an essential constituent in the makeup of smart communities. Unfortunately, personal information, explicitly submitted or implicitly inferable from observed behavior, poses evident privacy risks, especially when combined across several of the information sources mentioned in this work, and when enriched with metadata indicating size, location, time, frequency, and other contextual information.

The complex data collection process in the INRISCO system carefully addresses this challenge by means of a privacy layer composed of two main protection facilities, namely sanitization and statistical disclosure control, discussed next.

\subsubsection{Pseudonymisation, Redaction and Sanitization}
\label{sec:priv:1.1.1}

\noindent
The first privacy facility provided by INRISCO enables us to protect sensitive portions of the data collected by an automated process of textual suppression, called \emph{redaction}, or through categorical generalization, a mechanism termed \emph{sanitization}. Concisely, the internal architecture of this protection function contains three main operation parameters:
\begin{itemize}
	\item The transformation function provides any and all forms of \emph{anonymization}, or more generally, sanitization, including total and partial suppression, categorical and numerical generalization, and various forms of \emph{pseudonymization}.
	\item The pattern specification may be a choice of a predefined pattern, say an IP or email address, or a completely general regular expression to be matched and subsequently anonymized.
	\item Lastly, the search contextualization offers the possibility of restricting the search for an item, from the entire collection of objects to only a desired type.
\end{itemize}

\begin{figure}[htb]
	\centering
	\includegraphics[scale=0.9375]{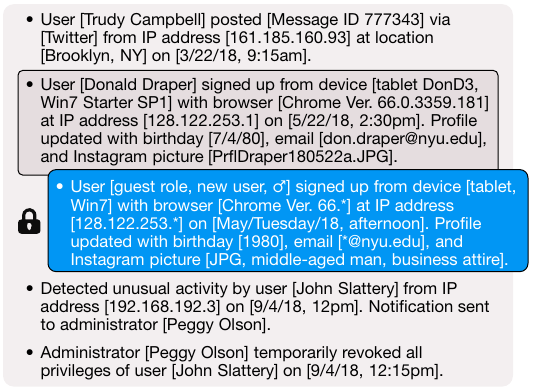}
	\caption{Hypothetical example of sanitization in semistructured data in INRISCO.}
	\label{fig:priv:1}
\end{figure}

A hypothetical example of sanitization of some of the semistructured data conceivably collected by our platform is illustrated in Fig. \ref{fig:priv:1}.

The automation of this process may rely on search and replacement patterns, flexibly expressed by means of regular expressions~\cite{sweeney96amia}, or resort to statistical frequency and information-theoretic properties of the items deemed sensitive~\cite{sanchez13ifs, sanchez15asist}. We would like to remark that sanitization may not only be applied to the functional data of a system, but also to cybersecurity data employed in the development of intelligent subsystems for automated threat detection~\cite{johnson16r, cipsec16prj}.

\subsubsection{Statistical Disclosure Control and $k$-Anonymous Microaggregation}
\label{sec:priv:1.1.2}

\noindent
Limited release or widespread publication of the incredible wealth of data collected by the INRISCO system would prove invaluable for a number of statistical studies aimed toward the management of smart communities. The second privacy facility offered by our privacy layer addresses the risks incurred by such release.

It was famously shown in \cite{sweeney00tr} that 87\% of the population in the U.S. might be unequivocally identified solely by the demographic triple consisting of their date of birth, gender, and 5-digit ZIP code. This striking finding, based on 1990 census data, holds in spite of the fact that in that year, the U.S. had a population of over 248 million. Thus, the study dramatically illustrates the discriminative potential of the simultaneous combination of a few demographic attributes which, considered individually, would hardly pose a real anonymity risk. It follows, ultimately, that the mere elimination of identifiers such as first and last name, or social security number, is grossly insufficient when it comes to effectively protecting the anonymity of the participants of released data containing confidential attributes linked to demographic variables. For that reason, publicly available attributes that narrow the identity of an individual or organization, typically attributes of demographic nature, are called \emph{quasi-identifiers}.

\begin{figure*}[htbp]
	\centering
	\includegraphics[scale=1]{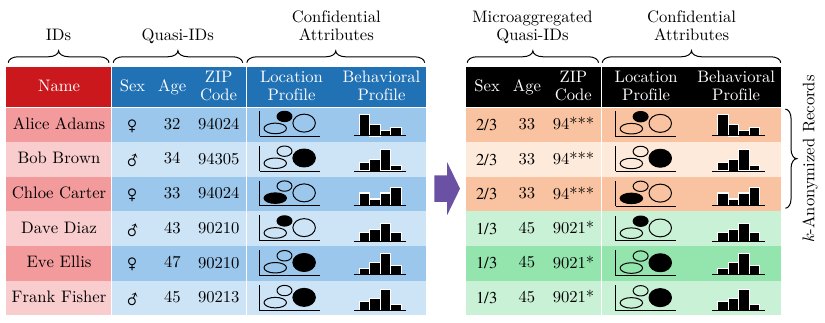}
	\caption{Synthetic example of $k$-anonymous microaggregation in INRISCO.}
	\label{fig:priv:2}
\end{figure*}

\emph{Statistical disclosure control} (SDC) concerns the postprocessing of the demographic portion of datasets containing sensitive personal information, in order to effectively safeguard the anonymity of the participating respondents. Intuitively, the perturbation of numerical or categorical quasi-identifiers enables us to preserve \emph{privacy} to a certain extent, at the cost of losing some of the \emph{data utility}, in the sense of accuracy with respect to the unperturbed version. $k$-\emph{Anonymity} is the requirement that each tuple of key-attribute values be identically shared by at least $k$ records in the dataset. This may be achieved through the \emph{microaggregation} approach illustrated by the synthetic example depicted in Fig. \ref{fig:priv:2}, where gender, age and ZIP code are regarded as quasi-identifiers, and location and behavioral profile as confidential attributes.

We must stress that the preservation of information utility is critical in the efficient management of some of the aspects involved in smart communities, particularly those related to emergencies and concerning the citizens' health. This is one of the reasons why this work advocates in favor of microaggregation over \emph{differential privacy}. While the former approach has been recently shown to carefully preserve the inference of macrotrends in machine learning~\cite{rodriguez18acc}, the latter method scrupulously observes privacy at the expense of considerable utility loss.

Because much of the data flowing through the INRISCO system is collected in real time, and might conveniently be made available for release with the lowest delay possible, conventional $k$-anonymous microaggregation is not perfectly suitable. Instead, we propose the $p$-probabilistic variant developed in~\cite{rebollo17ins}. Succinctly, $p$-probabilistic $k$-anonymous microaggregation resorts to a statistical model of respondent participation in order to aggregate quasi-identifiers in such a manner that $k$-anonymity is concordantly enforced with a parametric probabilistic guarantee. Owing to the possibility that some respondents may not finally participate, sufficiently larger cells are created striving to satisfy $k$-anonymity with probability at least $p$. The microaggregation function is designed before the respondents submit their confidential data. More precisely, a specification of the function is sent to them, which they may verify and apply to their quasi-identifying demographic variables, prior to submitting the micro-aggregated data along with the confidential attributes to an authorized repository.

\section{Sensing and Actuating Layer}
\label{sec:Sensing_Actuating}
\noindent
The {\em Sensing and Actuating} INRISCO layer depicted in Fig.~\ref{fig:INRISCO_architecture} is responsible of the dissemination of messages regarding a detected incident. In this section we detail the INRISCO operation to efficiently spread those warning messages. For this kind of environments, INRISCO mainly uses two kind of technologies. On the one hand, INRISCO disseminates warning messages through VANETs. Road safety applications envisaged for VANETs depend largely on the dissemination of warning messages to deliver information to concerned vehicles. The inherent VANET characteristics, make data dissemination an essential service and a challenging task in this kind of networks.  On the other hand, we propose to use Bluetooth low energy (BLE) and ad hoc Wi-Fi direct solutions. These two options will also work cooperatively and taking one or another will be decided in terms of distance and density of devices.

\subsection{Obtaining information from vehicles}
\label{sec:INRISCO_VANET_UPC}
\noindent

After the detection of an incident (e.g., traffic jam, car collision...), a vehicle could make a light and short video of the situation and send it through the VANET till reaching the closest road side unit (RSU) in the infrastructure of the city to alert the emergencies service (e.g., 911 or 112). Thus, vehicles participate in reporting a situation in the city using the VANET to have a quick reaction of the emergency and traffic management units in the city. With a video message, the level of seriousness of the incident could be better interpreted by the authorities than with a simple text message. Upon the arrival of such a video warning message, the INRISCO framework will check its veracity (see section \ref{sec:DBSCAN_UVIGO}). Afterwards, the emergency units will act consequently.

 The design of an efficient routing protocol to transmit video in VANETs is challenging due the inherent issues of VANETs (e.g., high speed of nodes, dynamic and heterogeneous network topology) and the resources demanded by video services. To contribute with this goal, we proposed in \cite{TVT_Ahmad_2017} a multimedia multimetric map-aware routing protocol (3MRP) used by vehicles to report video messages to RSUs through VANETs in smart cities. Our proposal includes five metrics to optimize the selection of the best next forwarding vehicle. Vehicles periodically broadcast the five metrics in their hello messages within their transmission range. Those metrics are:
\begin{itemize}
	\item  Distance to destination: It is the distance between each candidate vehicle and destination, i.e. the closest RSU.
	\item  Vehicles' density: It is computed as the number of vehicles in the neighbors' list of each node at the moment of sending the current hello message, divided by the area within the transmission range of that vehicle.	
	\item Trajectory: It is computed as a comparison of the current and future distances between a candidate vehicle to destination. This way we detect if the candidate vehicle is getting closer to or going away from  destination. The aim of this metric is to avoid that the source node could take wrong forwarding decisions based only on the distance and send packets to vehicles that were actually going away from destination, which could make packet losses increase as a consequence.
	\item  Available bandwidth estimation (ABE): ABE~\cite{carobw} is used to estimate the available bandwidth in the link formed between the current forwarding vehicle of the packet and each next hop candidate vehicle.
	\item  MAC layer losses: 3MRP uses the packet losses computed at the MAC layer in the link formed between the current forwarding vehicle and each candidate vehicle within its transmission range.
\end{itemize}

Then, the vehicle weights those five metrics using our dynamic self-configured weights (DSW) algorithm (in our previous work \cite{TVT_Ahmad_2017}). The goal is to obtain a multimetric score in order to arrange the nodes included in its neighbors' list and be able to select the best next forwarding node. This operation is repeated by each intermediate vehicle till the packet reaches its destination. The weights of the metrics are updated depending on the current state of the VANET. This way, we can highlight those decisive metrics that can better help the current forwarding vehicle to choose the best next forwarding node among the nodes in its neighbors' list. Further details of our 3MRP proposal are available in ~\cite{TVT_Ahmad_2017}. We implemented 3MRP in the NS-2 ~\cite{NS2} simulator to conduct a performance evaluation. The scenario is a 1700 {$\times$ 580 m$^2$} urban area obtained from OpenStreetMap (OSM)~\cite{OSM} and equivalent to Leganés, Spain (see Fig. \ref{fig:INRISCO_scenario}). SUMO~\cite{sumo} was used to generate the vehicular movement traces. All the figures show confidence intervals (CI) of 90 percent obtained from 20 simulations per point, each simulation with an independent mobility scenario. Vehicles send video reporting traffic information (MPEG2-VBR at 150 Kbps) towards the closest RSU. We analyzed the performance of 3MRP operating with the DSW algorithm (3MRP+DSW) to dynamically adjust the weights of the metrics. We compared the 3MRP+DSW results to 3MRP with fixed equal weights (3MRP), also to the well-known GPSR~\cite{GPSR} and finally to another similar proposal that also uses several metrics and transmits video over VANETs, named VIRTUS~\cite{VIRTUS}. We show results for low (50 vehicles/km{$^2$}) and medium (100 vehicles/km{$^2$}) vehicles' densities.

\begin{figure}
	\includegraphics[width=0.6\textwidth]{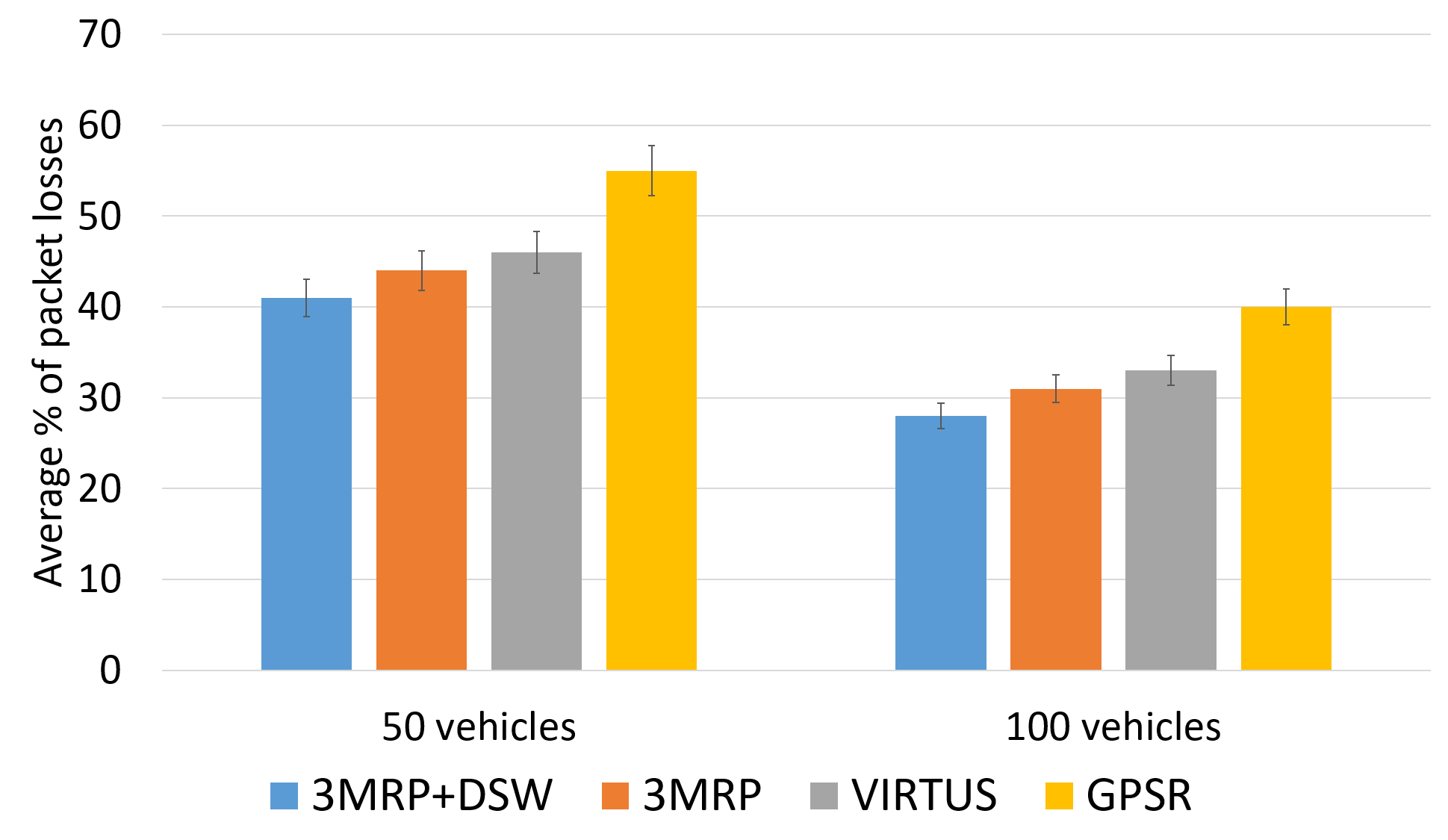}
	\caption{Average percentage of packet losses. }
	\label{fig:losses_Ahmad}
\end{figure}

\begin{figure}
	\includegraphics[width=0.6\textwidth]{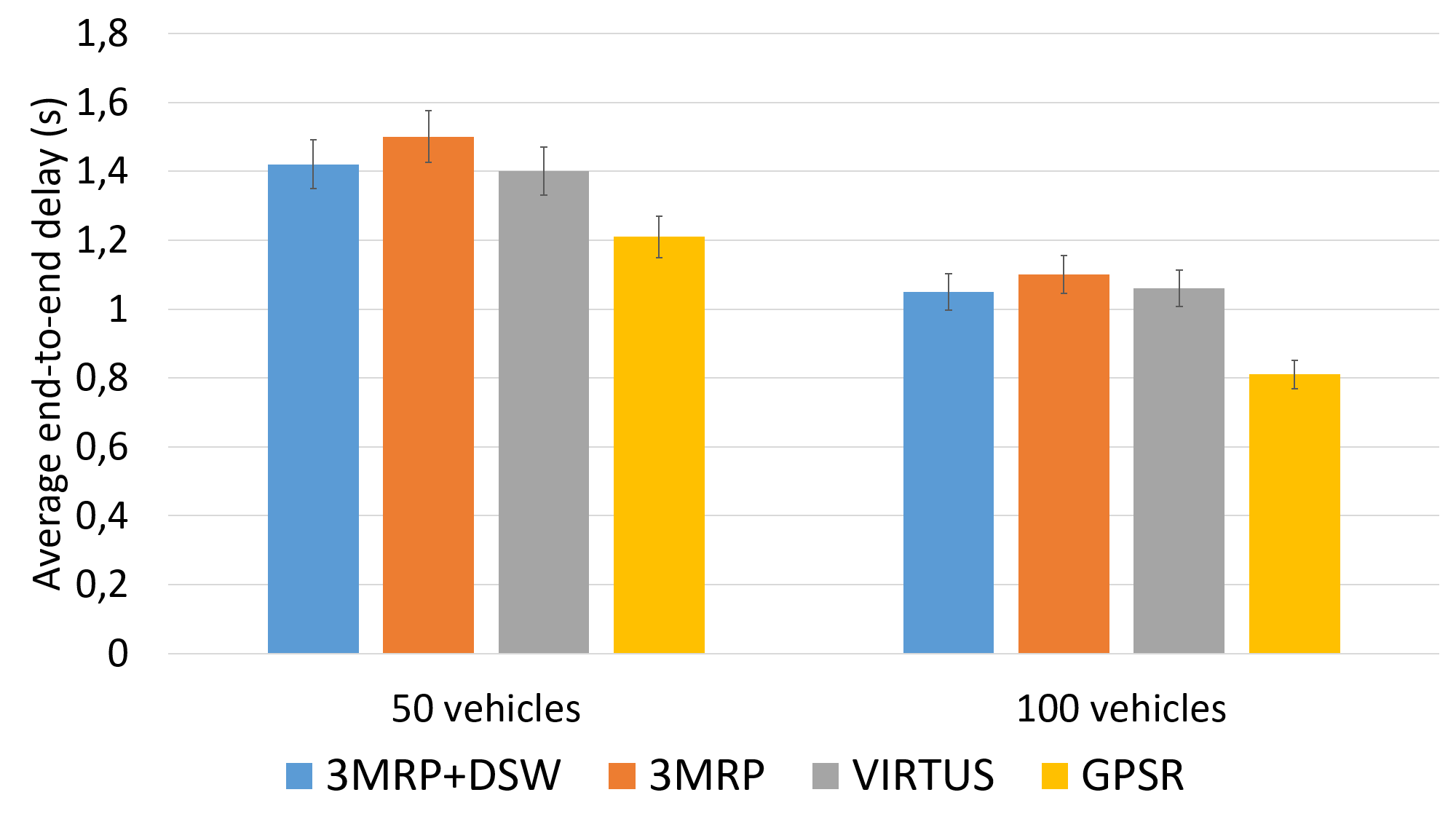}
	\caption{Average end-to-end packet delay. }
	\label{fig:delay_Ahmad}
\end{figure}

\begin{figure}
	\includegraphics[width=0.6\textwidth]{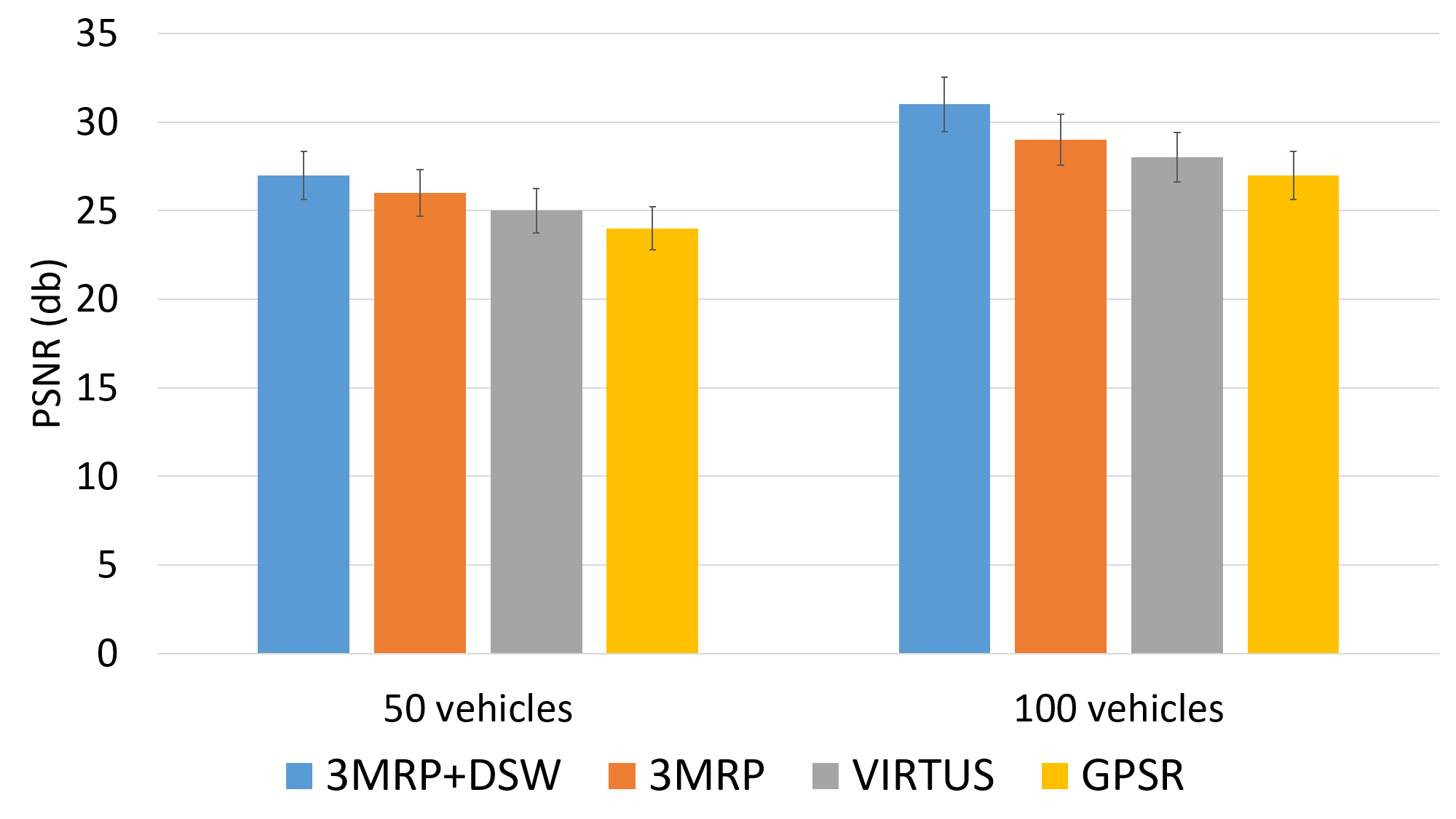}
	\caption{Peak signal to noise ratio (PSNR). }
	\label{fig:PSNR_Ahmad}
\end{figure}

In Fig.~\ref{fig:losses_Ahmad} we can see that 3MRP+DSW obtains the best results for both vehicles' densities reducing losses around 10\% with respect to GPSR and 6\% with respect to VIRTUS. This is due to the optimal selection of the next forwarding node based on the five proposed metrics with dynamic self-configured weights. Fig.~\ref{fig:delay_Ahmad} shows the average packet delay, which is calculated for those packets that successfully arrived at destination. Since GPSR takes forwarding decisions considering only distance, it obtains the lowest packet delay in both scenarios. However, GPSR shows the highest losses (see Fig.~\ref{fig:losses_Ahmad}). That is, with GPSR a lower number of packets arrived at destination but through shorter paths so that latency decreases. Conversely, 3MRP reduces packet losses in 15\%, whereas the delay slightly increases in about 0.3 seconds, compared to GPSR. Fig.~\ref{fig:PSNR_Ahmad} depicts the peak signal-to-noise ratio (PSNR). Both 3MRP versions outperform GPSR and VIRTUS in 4 dB and 1 dB, respectively. 3MRP+DSW improves the PSNR in 2 dB compared to 3MRP, so using dynamic weights further improves the performance since it classifies nodes in a better way giving each metric its importance depending on the current environment conditions.

\subsection{Efficient dissemination of warnings and advertisements in VANETs considering intersections}
\label{sec:Sensing_Actuating_UC3M}
\noindent
When an abnormal situation is detected in a smart city, VANETs can help sending relevant information (such as blocked road warnings) to all vehicles in a region of interest. This multi-hop dissemination of information to all vehicles present particular features and difficulties that may require special measures. The dissemination of information may cause broadcast storms, thus the design of an efficient smart dissemination algorithm is of paramount importance. Urban scenarios are especially sensitive to broadcast storms because of the potential high density of vehicles in downtown areas. They also present numerous crossroads and signal blocking due to buildings, which make dissemination more difficult than in open and almost straight interurban roadways.

When the message is extremely critical (e.g., a traffic accident), broadcast storms are a cost we could afford. Nonetheless, for many other not so critical warnings or advertisements, alternative ways to efficiently disseminate this information are needed. In the INRISCO project, we have discussed several options to avoid the broadcast storm problem while trying to achieve the maximum coverage of the region of interest.

\begin{figure} [H]
	\centering
	\includegraphics[width=0.6\textwidth]{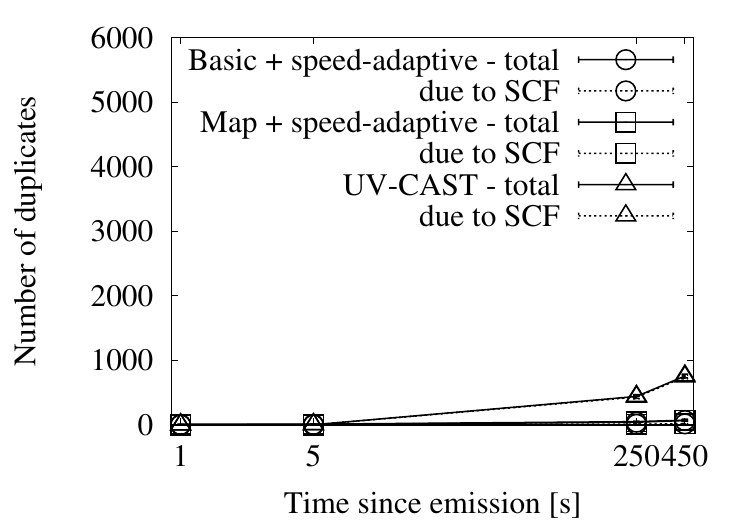}
	\caption{Number of duplicates sent for adaptive speed and map approaches, with 25 vehicles/km$^{2}$ in our urban scenario.}
	\label{fig:messages-evol-25}
\end{figure}

\begin{figure} [H]
	\centering
	\includegraphics[width=0.6\linewidth]{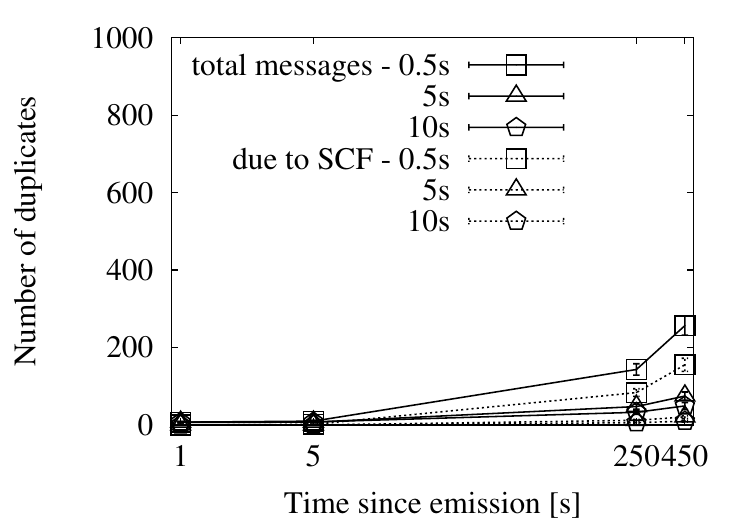}
	\caption{Number of duplicates sent for the fixed interval approach, with 25 vehicles/km$^{2}$ in our urban scenario.}
	\label{fig:13}
\end{figure}

In \cite{SENSORS2016} we evaluate different ways to detect and take advantage of intersections. Broadcast storms are usually avoided by using broadcast suppression schemes. They consist of a set of rules that prevent some nodes from retransmitting the message. In order to combat network partitioning, a popular type of mechanism is store-carry-forward (SCF), where one or more vehicles get in charge of keeping a copy of the message and forward it after some time (for example, when they encounter other vehicles). Our conclusion is that we can apply two different schemes depending on the requirements of the application. If the source will repeat the message periodically, then using store-carry-forward does not make sense, and the map-based urban adaptation (i.e., being aware of intersections) should be applied. Otherwise, it is a good idea to use store-carry-forward. We added a mechanism that let vehicles detect when they are passing through an intersection and increase their chance to forward messages if so. This way, we increase the chances of the message to be disseminated in other directions and therefore reach more vehicles. The timer for retransmissions can be set in different ways and we have considered three schemes: (i) with a fixed interval, (ii) based on the current vehicle's speed (i.e., the faster the vehicle, the shorter the retransmission interval), and (iii) by polling the position of the vehicle in a digital map to check if it is at an intersection (i.e., forwarding when the vehicle is located at an intersection). The performance results we have obtained show that speed-adaptive approaches offer a better compromise on coverage and overhead.
In Figures~\ref{fig:messages-evol-25} and~\ref{fig:13} we can see the number of duplicate messages sent when using different values of fixed intervals (Figure~\ref{fig:13}) and using speed-adaptive and map-based strategies (Figure~\ref{fig:messages-evol-25}).

Results for another state of the art approach  ~\cite{UV-CAST} called urban vehicular broadcast protocol (UV-CAST) are also shown for comparison. UV-CAST is a broadcast scheme for VANETs in urban environments. It is intended to work in either well-connected or disconnected vehicular networks without help from fixed infrastructure. Depending on the received beacons, a vehicle can determine if it is inside a connected region or not. If it is, it will act according to the well-connected regime, applying a broadcast suppression scheme. Otherwise, being it a boundary node or a totally disconnected node, it will enter the disconnected regime and store-carry-forward the message.

Figure~\ref{fig:messages-evol-25} shows that both the map-polling and the speed-adaptive approaches of our scheme are good, though the latter offers a better compromise on coverage and overhead. When compared to UV-CAST, we achieve a better coverage in the first stages with either of the two options. UV-CAST gets an almost total coverage later. Nevertheless, as figures show this is at the cost of the high amount of duplicates that the UV-CAST store-carry-forward mechanism produces.

\subsection{Efficient game-theoretical dissemination of warning messages in urban VANETs}
\label{sec:Sensing_Actuating_UPC}
\noindent
Safety applications for VANETs normally rely on broadcast-based protocols, which have the task of disseminating emergence messages quickly and efficiently through the network. Thus, a key research problem is the design of a scalable information dissemination algorithm that can efficiently work with high reliability and short delay under different network conditions.
To tackle this issue, it is required to solve the broadcast storm and consider the intermittently connected network. Inherent features of VANETs that we need to take into account are high mobility of vehicles, unstable link connectivity, fading of signal, and obstacle-constrained environments.

Beaconing or one-hop broadcast is one of the most common techniques for data dissemination in VANETs. Exchanging beacon messages is important for road safety applications. However, channel load may increase too much in scenarios with high vehicle density when the beacon rate is fixed. Within this mindset and assuming decentralized knowledge of the topology, efficient broadcasting could be formulated as a classical constrained optimization problem with the objective of minimizing the number of re-transmissions while at the same time guaranteeing high delivery ratio.

In~\cite{Iza18}, we presented an adaptive distributed dissemination (ADD) protocol for data dissemination in VANETs. ADD was designed to operate without any roadside infrastructure in urban scenarios under diverse road traffic conditions. ADD uses a decentralized stochastic solution for the broadcast data dissemination problem through two game-theoretical mechanisms. Game theory can be used to design mechanisms to predict behavior in situations where a state is the result of a series of interactions between different nodes (referred as \textit{players} in the game), who act according to their own preferences regarding future performance and existing incentives.

Firstly, the {\em asymmetric volunteers dilemma game} ~\cite{Diekmann1993} was adapted to VANETs as a mechanism to cope with the broadcast storm problem. The probability that a node forwards a broadcast message was computed using the number of candidate vehicles to forward the message, i.e. the number of vehicles that are listening to the transmission. In~\cite{Iza18} we designed a utility function used by vehicles to decide to forward or not the received message. Our proposal of utility function (see Eq. ~\eqref{eq:Utility_Function}) includes information about the local neighborhood of a vehicle $i$ by means of two metrics: (i) distance (between the vehicles that are source of the warning message and the current receiver) factor ($Df_i~$) and (ii) link quality (measured from a combination of signal quality, channel quality and collision probability) factor ($LQf_i$). Power of ten is used to be more sensitive to changes in the environmental conditions. The weights $\alpha_1=6$ and $\alpha_2=4$ were obtained through extensive simulations under different traffic conditions, Further details are available in~\cite{Iza18}.

\begin{equation} \label{eq:Utility_Function}
U_i(Df_i,LQf_i)  =10^{(10\, -\, ( \alpha_1 \cdot Df_i \, +\, \alpha_2 \cdot LQf_i))}
\end{equation}

Regarding the distance factor $\mathrm{\textit{Df}_\emph{i}}$, if the receiving vehicle is not located over an intersection, then $\mathrm{\textit{Df}_\emph{i}}$ is calculated as the ratio between $\mathrm{\textit{D}_{\textit{sr}}}$ and the transmission range $\mathrm{\textit{R}_{max}}$. With this, the farthest vehicle from the sender is assigned the highest distance factor. 

On the other hand, if the receiving vehicle is located over an intersection, the distance factor $\textit{Df}_i$ is calculated as a decreasing function of the distance $\mathrm{\textit{D}_{\textit{rint}}}$. This way, the lower $\mathrm{\textit{D}_{\textit{rint}}}$, the~higher its $\textit{Df}_i$~which means that the vehicle is a good candidate to broadcast the message. The distance factor is summarized in Eqs.~\eqref{eq:df1}.

\begin{equation}
\textit{Df}_i= \frac{D_{sr}}{R_{max}} \ \text{, if } D_{r\textit{int}}>R_{max} ; \qquad \textit{Df}_i= \text{1}-\frac{D_{r\textit{int}}}{D_{r\textit{int}}+1}  \ \text{, otherwise}
\label{eq:df1}
\end{equation}

\begin{figure}
	\includegraphics[width=0.6\linewidth]{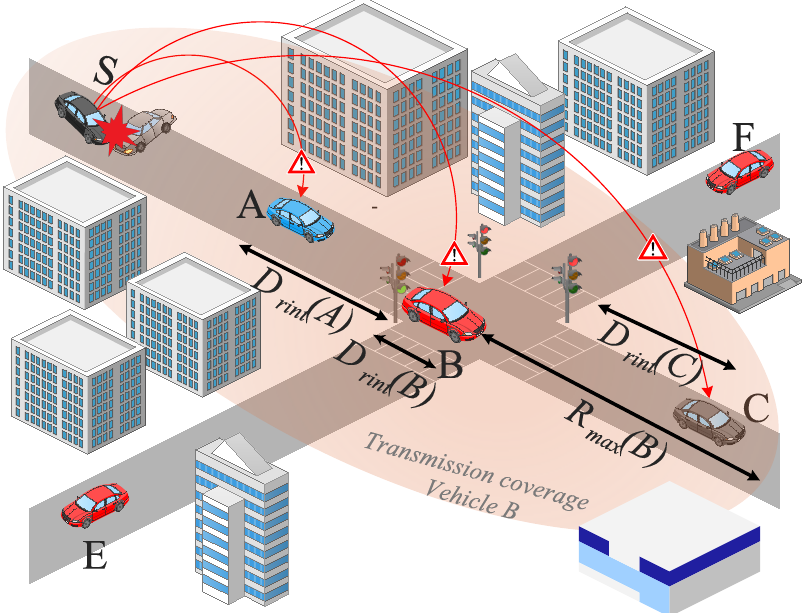}
	\caption{Distance factor $Df_i$. Case with $D_{rint}\leqslant{R}_{max}$.}
	\label{fig:intersectionMinor}
\end{figure}

Figure~\ref{fig:intersectionMinor} presents the scenario when the vehicles receiving the message have intersections within their transmission range. In this case, vehicles $A$, $B$, and $C$ compute the distance factor according to  Equation~\eqref{eq:df1}. Hence, vehicle B that is crossing the intersection is assigned the highest distance factor. We refer the reader to ~\cite{Iza18} where the rest of detailed explanations are available.

Secondly, the {\em forwarding game} ~\cite{NaserianTepe2014} was evaluated as another mechanism to mitigate the broadcast storm problem. In this case, the strategy of the players is to select a forwarding probability that maximizes their pay off using a utility function designed as a function of the {\em player's availability} and the forwarding probability of other players. 

Availability of a player is a normalized factor based on metrics such as distance from the source of the warning message (e.g., a crashed vehicle) and estimated bandwidth of the link formed between the node currently holding the packet and each candidate node (to forward that message) within its transmission range. The utility function designed for this case can be found in~\cite{Iza18}. Thus, either the asymmetric volunteers dilemma game or the forwarding game are played whenever vehicles receive a broadcast message and they choose in a decentralized way their best strategy to broadcast the message or not. That is, from the utility function vehicles compute the probability to forward or not the received message, as it is explained in ~\cite{Iza18}.

Because exchanging vehicle information via beacon messages is important for active safety applications, our proposal includes an adaptive traffic beacon (ATB) protocol~\cite{Sommer2011a} in order to purchase an uncongested channel, trying to prevent packet losses due to collisions and to reduce the end-to-end delay of the information transfer. Also, our proposal includes a store-carry-forward mechanism to relieve the intermittently connected network problem presented on streets or roads that have low-density traffic conditions.

To carry out the performance evaluation we employed OMNeT$++$~\cite{Varga10} together with VEINS~\cite{Veins2015, sommer2011bidirectionally} as the event-based network simulator for VANETs, and SUMO to generate the vehicular movement traces \cite{sumo}. In addition, to dispose a realistic scenario we used real maps from OpenStreetMap~\cite{OSM}. Specifically, in this work we used a map for a scenario equivalent to the one that corresponds to a 2.5x2.5 km$^{2}$ urban area in Legan\'es, Spain (see Fig. \ref{fig:INRISCO_scenario}). In these simulations, a vehicle disseminates a video warning message encoded with H.265/HEVC ~\cite{hevc} and publicly available at ~\cite{trace2017}.

\begin{figure} [htpb]
	\centering
	\includegraphics[width=0.6\textwidth]{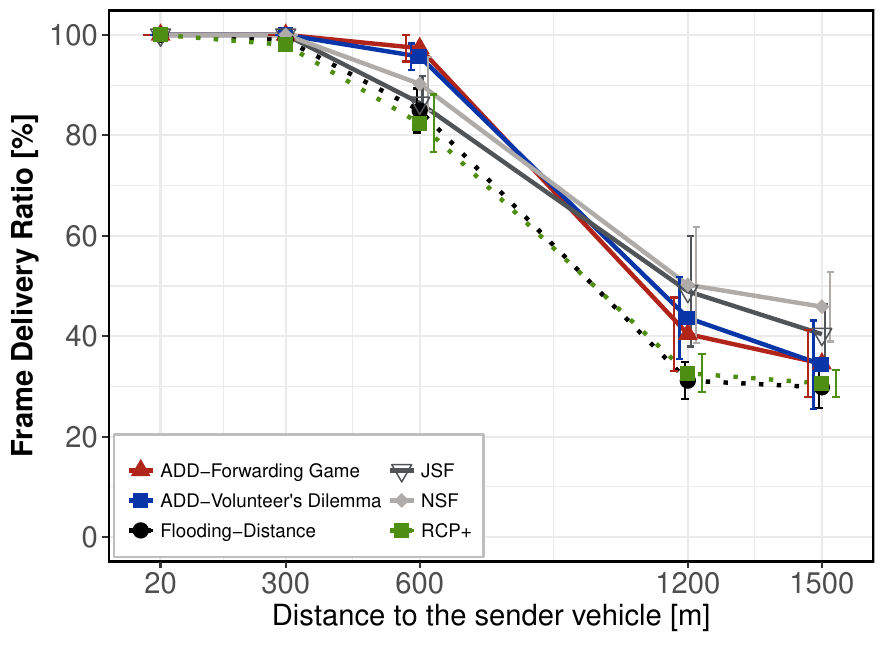}
	\caption{Frame delivery ratio (FDR) for 40 vehicles/km$^{2}$.}
	\label{fig:frameDeliveryRatio40Urban}
\end{figure}

\begin{figure} [htpb]
	\centering
	\includegraphics[width=0.6\textwidth]{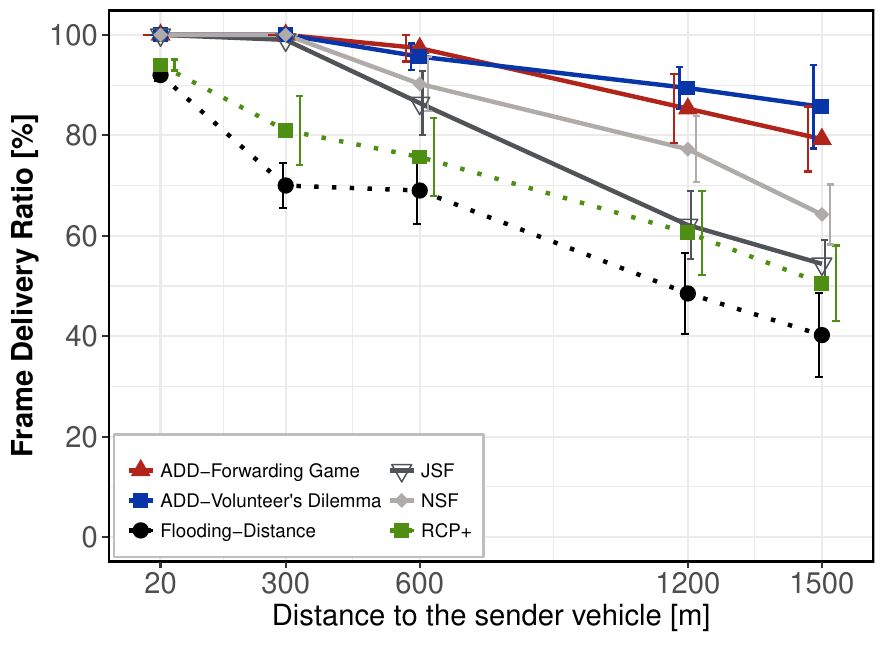}
	\caption{Frame delivery ratio (FDR) for 100 vehicles/km$^{2}$.}
	\label{fig:frameDeliveryRatio100Urban}
\end{figure}

Figures~\ref{fig:frameDeliveryRatio40Urban} and \ref{fig:frameDeliveryRatio100Urban} show the frame delivery ratio (FDR) for vehicles' densities of 40\% and 100\%, respectively. Figures show 95\% confidence intervals for 10 repetitions per point with independent seeds. Our AAD proposal used with the two game-theoretical algorithms adapted for smart dissemination in VANETs (AAD\_Asymmetric and AAD\_ForwarGame in Figure~\ref{fig:frameDeliveryRatio40Urban}) shows benefits compared to other similar state-of-the-art dissemination protocols: Junction Store and Forward (JSF)~\cite{Sanguesa2014}, Neighbor Store and Forward (NSF)~\cite{Sanguesa2014a}, distance-flooding~\cite{TorresPinolCalafateEtAl2014}, and our previous proposal RCP+~\cite{Iza-ParedesMezherAguilarIgartua2016a}.

Figure~\ref{fig:frameDeliveryRatio40Urban} shows the FDR for a light vehicle density of $40$ veh./km$^{2}$. Low density of vehicles affects the ability of the protocols to disseminate through the VANET. In fact,~only~vehicles located up to~(300 m) from the accident received the complete trace. At 600 m from the accident ADD reaches a FDR of 97\% and 95\% with forwarding game and volunteer's dilemma, respectively. In the NSF, JSF, Flooding-Distance and RCP+ schemes we obtain an FDR of 90\%, 86\%, 84\% and 82\%, respectively. For vehicles located at 1200 and 1500 m all the protocols keep a FDR below 50\%. This has sense because in such a sparse scenario there might happen that no vehicle is in neighborhood to receive and disseminate the video packet to other vehicles around.

 Figure~\ref{fig:frameDeliveryRatio100Urban} shows the FDR for $100$ veh./km$^{2}$. Higher density of vehicles improves the FDR. This can be noticed at 600 m, 1200 m and 1500 m where the FDR increases with respect to Figure~\ref{fig:frameDeliveryRatio40Urban} for all the tested schemes. For instance, at 600 m from the accident, ADD~reaches an average maximum rate of 97\% and 95\% with forwarding game and volunteer's dilemma, respectively, while in the JSF, NSF, Flooding-Distance and RCP+ schemes, we~obtain an FDR of 86\%, 90\%, 69\% and 75\%, respectively.  At the RSUs located at 1500 m, JSF,~NSF, Flooding-Distance and RCP+ schemes keep a FDR below 64\% of received frames. Conversely, ADD~reaches an average maximum rate of 79\% and 85\% with forwarding game and volunteer's dilemma, respectively. Here, we can notice how the game-theoretical schemes allow us to achieve a~better performance in comparison to the other schemes. We refer the reader to ~\cite{Iza18} to see further details of our proposal AAD and its performance evaluation.

\subsection{Dissemination of alerts or recommendations using BLE and Wi-Fi}
\label{sec:Sensing_Actuating_UVIGO}
\noindent
Whereas the main goal of the INRISCO framework is the quick detection of incidents in cities, dissemination of alerts or recommendations for citizens to react properly to the unexpected identified events is the second pillar of the INRISCO system. Although those messages could be shared using social media, it seems more efficient to collect them from users that are actually in the incident area and share them using a proximity-based peer-to-peer communication technology. This increases trust and provides a real time source of information about nearby incidents (e.g., traffic jams or street works). 

\begin{figure}[htpb]
\centering\includegraphics[width=0.6\linewidth]{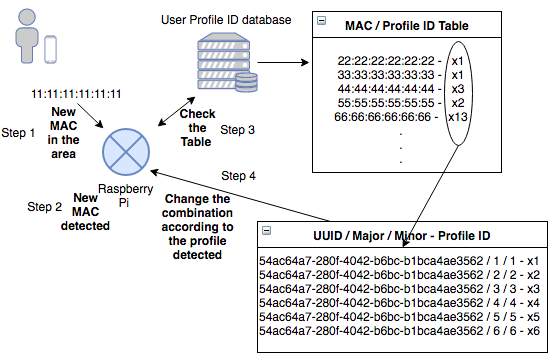}
\caption{BLE scheme to transform static ibeacons in dynamic ibeacons.}
\label{fig:MAC_profile}
\end{figure}

The BLE \cite{BLE} technology is a relatively new technology highly optimized for low energy networking operations, as the name suggests. As a low energy data broadcaster, it represents a potential technology for real-time advertisement and  message dissemination in nearby areas. One of the latest addition to this technology, Beacons, or to say iBeacons \cite{ibeacon}, is mostly used in a retail, health care and smart homes area. We propose a nearby  BLE ibeacons architecture \cite{Miran2017BLE} to support a dynamic interest-based system for broadcasting information. This dynamic information is categorized by different numbers, which are all part of iBeacon universally unique identifier (UUID) hierarchy architecture \cite{Miran2018BLE}. With this, we define a mechanism that supports mapping users' interest according to specific taxonomies, and these taxonomies are also matched to the UUID classification \cite{Miran2018journal}. The proposed architecture is shown in Figure \ref{fig:MAC_profile}, were the steps involving a new user that accesses the broadcasting area are detailed. First, the MAC address is obtained by the first beacon device (iBeacon advertiser) and verified to be in the database. If so, the configuration of the beacon
advertiser (Raspberry Pi in our implementation) is changed, and a new combination of UUID, Major, and the Minor number instructs to the user's device application to display corresponding message data.

In order to cover broader areas, INRISCO also uses ad hoc Wi-Fi (multi-hop messages) to supplement the BLE structure. Because of this distributed structure, contrary to the centralized approach in social media dissemination, with ad hoc Wi-Fi every device receives those warning messages, whether they are interested or not. Consequently, it does not add any benefit, specially in a crowded scenario where the network may even get congested and collapse. Ironically, in this case, too much information may prevent users from being informed. With the aim of solving this and to prevent unnecessary transmissions, we have also defined \cite{Castro-Jul2016} a collaboration scheme that involves neighbor nodes in the assessment of alert pertinence. Unlike other alert systems, this assessment is performed in a distributed manner without any centralized entity or hierarchical structure. Starting from this base, we have also incorporated a content-aware assessment mechanisms \cite{CASTROJUL2018129} to deal with a common problem in participatory sensing: information assessment. This approach establishes how users may contribute to sensing tasks and alerts generation in an efficient way. Encouraging users' participation may result in large amounts of data, which may not be valid or relevant. Our strategy prevents duplicates from congesting the network, but it does not asses every generated alert and does not deal with low-quality or irrelevant alerts. In order to ensure users receive only interesting alerts and the network is not compromised, we propose two collaborative alert assessment mechanisms that, while keeping the network flat, provide an effective message filter. Both of them rely on opportunistic collaboration with nearby peers. In fact, we have proposed two collaboration schemes. On the one hand, the CTD-query ({\em query-based collaboratively-triggered dissemination}), where the first node that detects the event sends a request to their neighbors that must decide whether to confirm the alert or not. Only if most of the received replies in a certain time are confirmed, the alert is broadcast. If not, it is discarded. On the other hand, the CTD-passive ({\em passive collaboratively-triggered dissemination}) proceeds as follows. Nodes send proposed alerts to their neighbors, which compare them with their own or previously received ones but do not elaborate any reply; so alerts are always proactively decided by every one of the nodes. Therefore, the impact of receiving irrelevant messages is minimized.  
 \begin{figure}[htbp]
 \centering
 \subfigure[200 devices \label{messages-decrease-100}]{\includegraphics[width=0.3\textwidth]{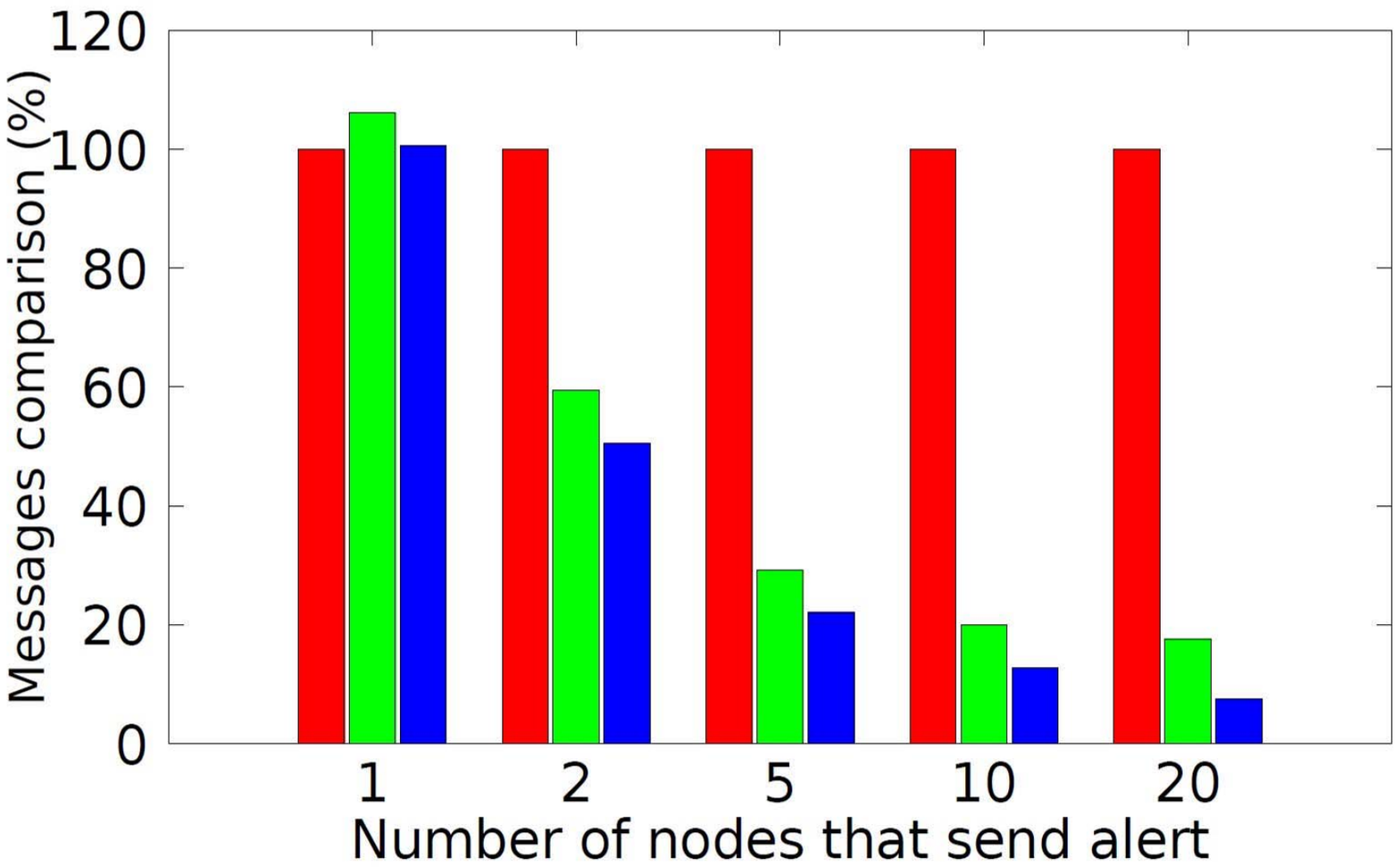}}
 \subfigure[500 devices \label{messages-decrease-1000}]{\includegraphics[width=0.3\textwidth]{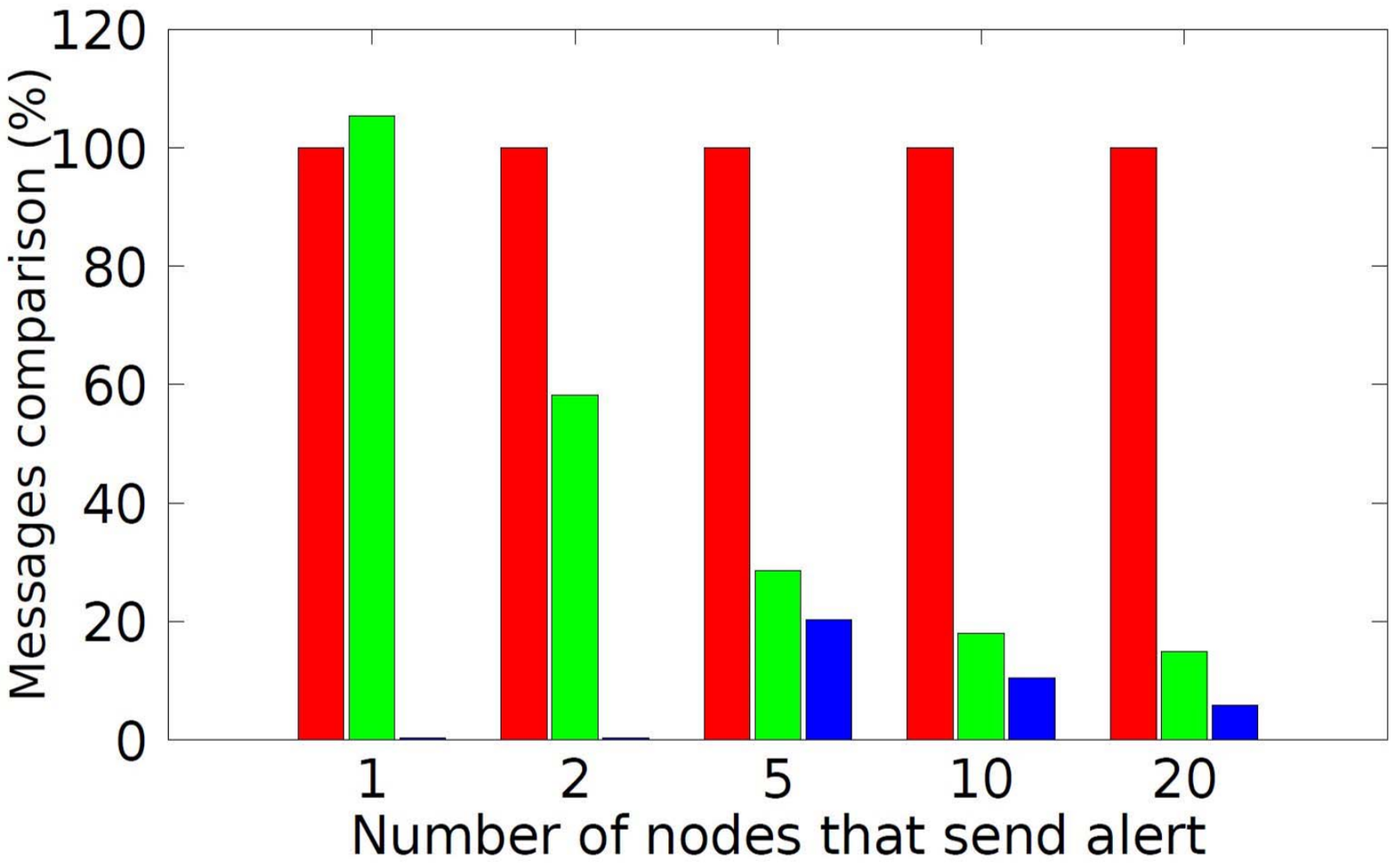}}
 \subfigure[1000 devices \label{messages-decrease-5000}]{\includegraphics[width=0.3\textwidth]{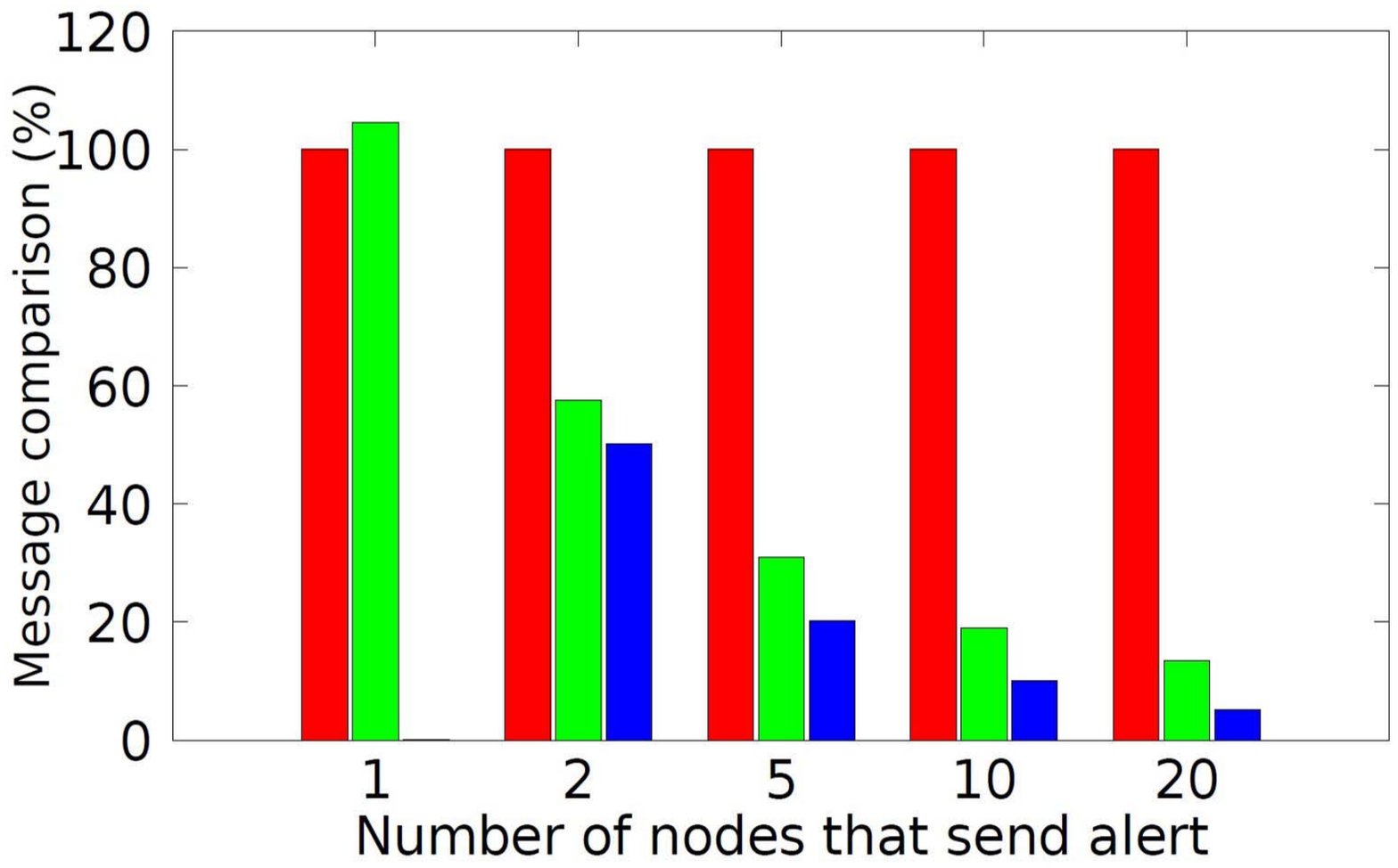}}
 \caption{Percentage of messages sent with CTD-query (green bar) and with CTD-passive (blue bar) in relation to the dissemination without assessment (red bar).}
 \label{fig:individual2}
 \end{figure}
 
\begin{figure}[!htbp]
 \centering
 \subfigure[200 devices \label{messages-100-prob}]{\includegraphics[width=0.30\textwidth]{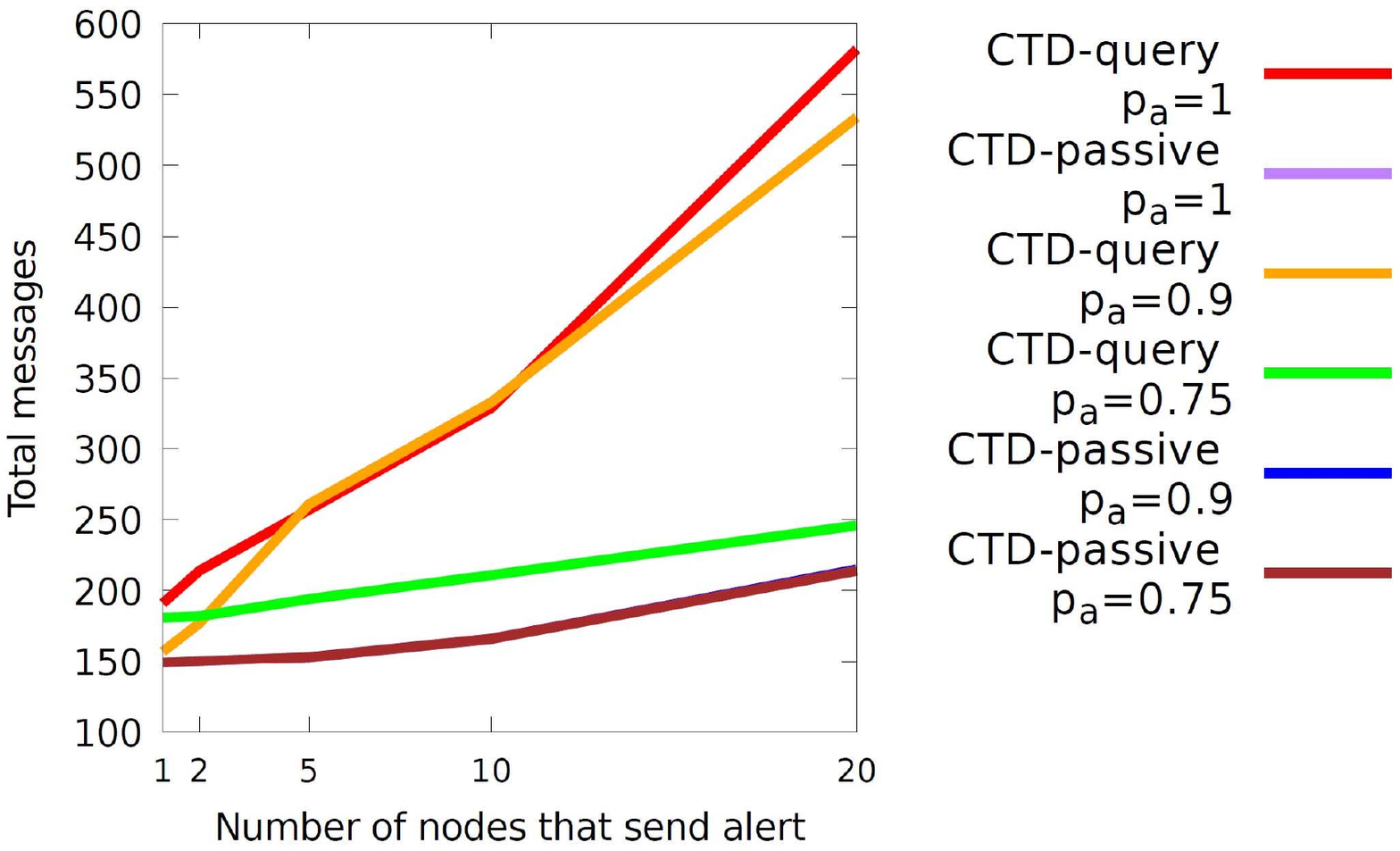}}
 \subfigure[500 devices \label{messages-500-prob}]{\includegraphics[width=0.30\textwidth]{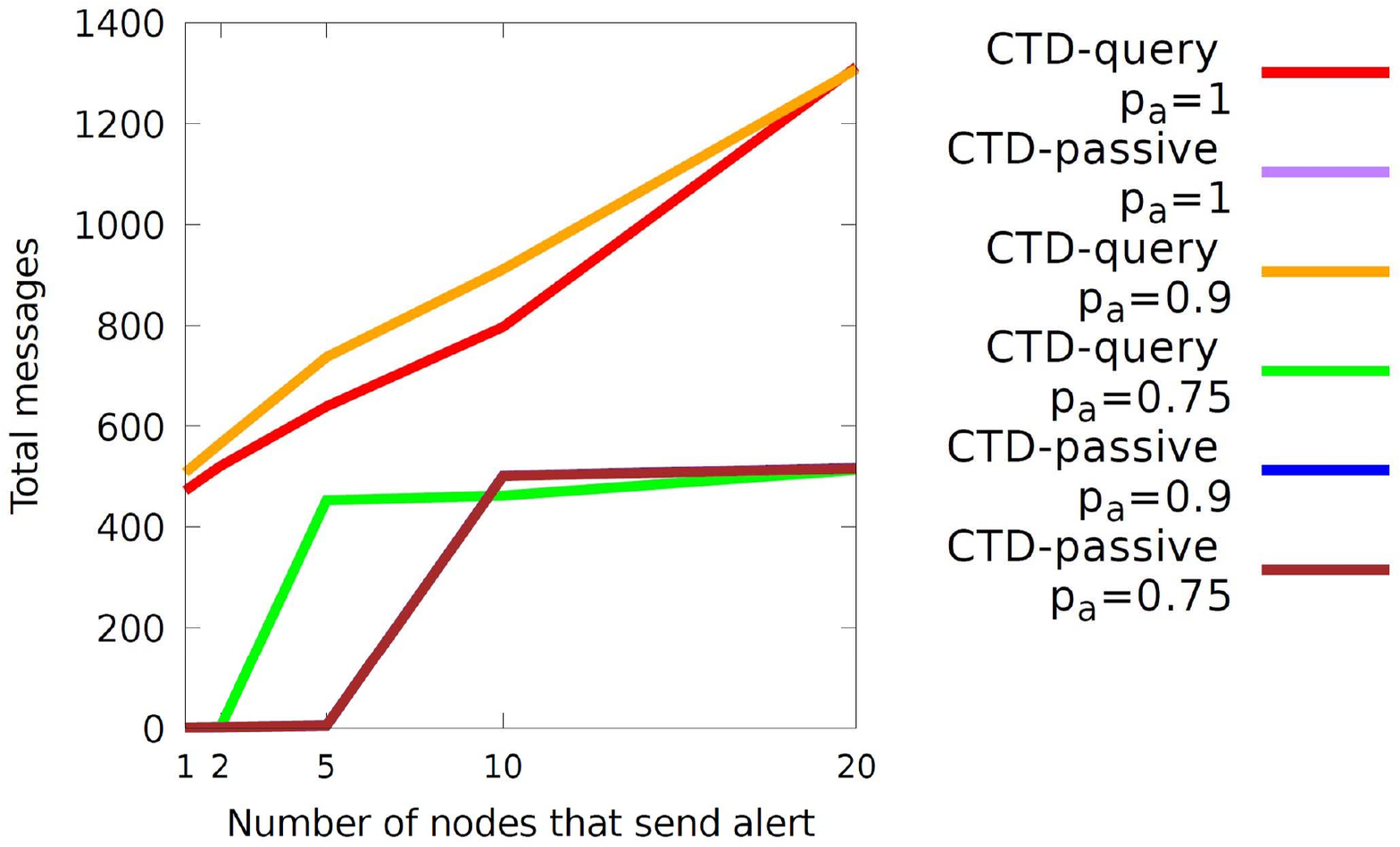}}
 \subfigure[1000 devices \label{messages-1000-prob}]{\includegraphics[width=0.30 \textwidth]{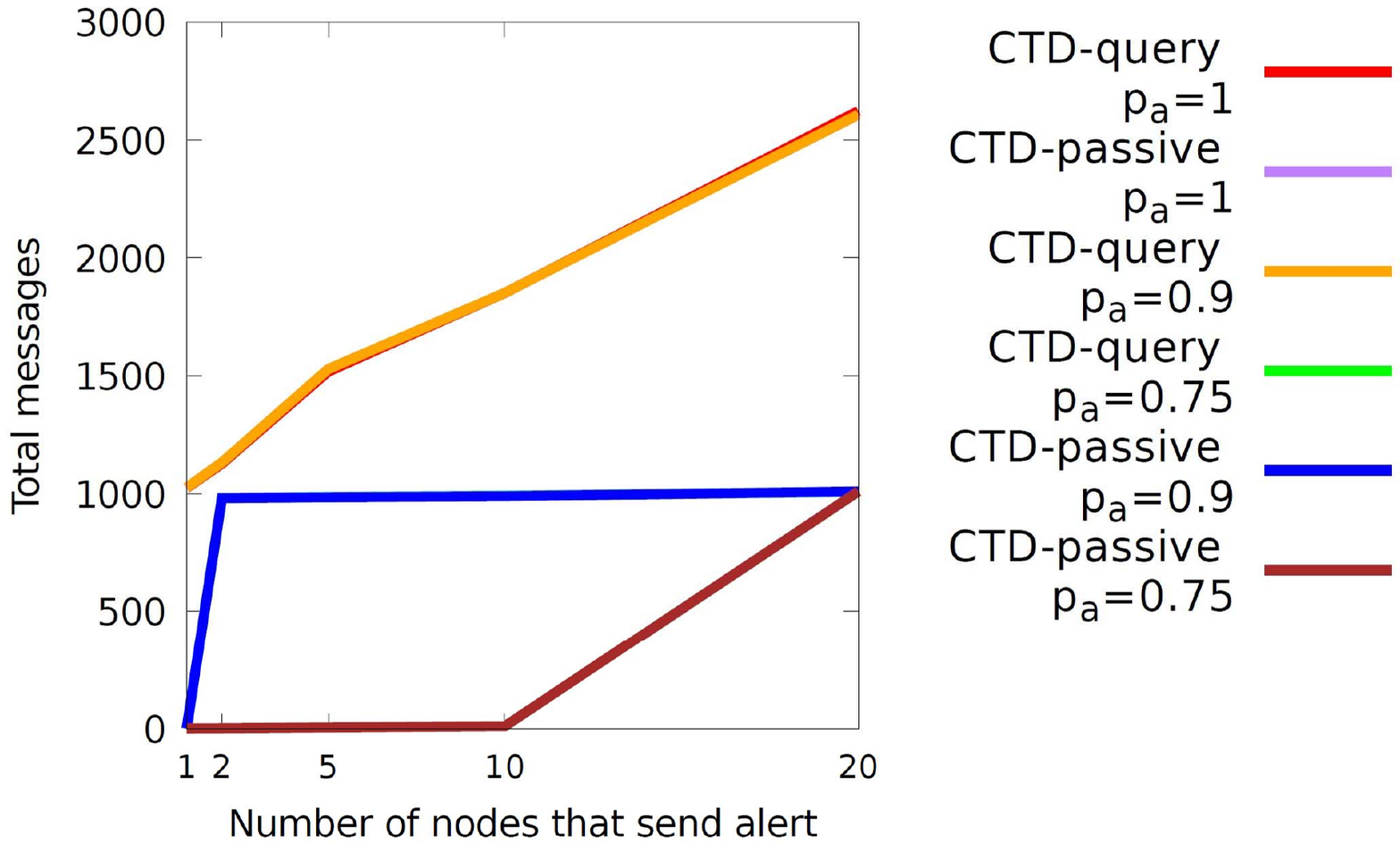}}
 \caption{Number of messages sent with CTD-query (green line) and with CTD-passive (blue line) compared to the dissemination without assessment (red line).}
 \label{fig:individual-prob}
 \end{figure}

In order to assess the proposal, we have carried out simulations using the following tools: OpenStreetMat \cite{OSM} to have the map of the area, SUMO urban mobility simulator \cite{Krajzewicz2012} to populate the area under study with random pedestrian routes and ns-3 network simulator \cite{ns3} where we have implemented the two strategies and carried out the simulations. In fact, we have compared what happens with 100, 200, 500, 1000 and 5000 pedestrians in the area with different device density. In each one of those, we have simulated situations where 1,2, 5, 10 or 20 nodes alert about the supposed incident. The comparison was based on different criteria: Message efficiency, delivery rate, transmission delay between opposite area edges and transmission time to cover the alert area. Figure \ref{fig:individual2} shows results regarding the percentage of messages sent compared to the dissemination without any assessment. According to the results, we can observe a clear trend on the system behavior (more remarkable as the number of devices increases) where for one alert sender the efficiency of the message dissemination with CTD-query (green bars) outperforms the message dissemination without assessment (red bars). The reason is that CTD-query does not require any extra messages apart from the dissemination ones. Figure \ref{fig:individual-prob} shows the number of messages as a function of the assessment probability, i.e. the probability that a node refuses to accept the alert. The figures show different results that variate with the value of $p_{a}$. Notice that in general CTD-passive involves less messages in total.  

Finally, and in line with this idea to certificate or assess the information given by peers, we have recently defined a certification architecture based on a fog approach \cite{Fatima2017UcaMI}, where neighbors are validated according to their location in a collaborative scheme to verify if they are trustworthy.

\section{Related Work}
\label{sec:related_work}
\noindent

In this section we highlight several research works related to our proposals. We have classified similar proposals in (i) those related to the detection of situations in urban environments, (ii) those regarding cellular-based traces, (iii) some proposals on VANETs to gather traffic information and disseminate warning messages, and (iv) those regarding data privacy and trust.

\subsection{Events detection in urban areas}
\noindent

Early detection of unusual crowds in urban areas is a challenge that has been tackled from different perspectives and using different technologies. Perhaps the most popular are those approaches based on the analysis of data gathered from video-surveillance systems, like in  \cite{ge2012vision}, where smalls groups of pedestrians who are walking together can be detected, or other proposals that focus on the detection of unexpected events using video-data, like in \cite{hamid2005detection} or \cite{xu2016detecting}.

However, the analysis of data gathered from LBSNs has become an interesting option, since they provide two kind of valuable information: the content of the posts (text, images, etc.) and the location from it was posted. Some interesting approaches focus on analyzing moving patterns of citizens, like in~\cite{CHEN2015156}, which focuses on trying to predict the most probable next location; in~\cite{shelton2015social}, which focuses on analyzing geotagged tweets to analyze the mobility between two specific areas in a specific city; or in~\cite{Trasarti2017350}, where the analysis is focused on profiling users based on their mobility patterns. 

On the other hand, the analysis of the content of the posts (mainly text) is a challenge that has been tackled to event detection in the literature \cite{weiler2016evaluation,Zhou:2014}. The authors of \cite{arin2018twec} propose a web service able to clustering tweets based on their semantic similarity. However, and apart from this approach, the location information linked to the shared posts has also brought several proposals to detect crowds and events in urban areas, like in \cite{ferrari2011extracting}, which analyzes the most visited locations in a city, or in \cite{chae2012spatiotemporal}, where the objective is finding out popular topics. Local events are also the focus in \cite{walther2013geo}, with an approach which constructs clusters of tweets according to their density. In \cite{ranneries2016wisdom} posts from both Twitter and Instagram are clustered according to their hashtags, and then geotagged to associate a single place to each cluster. Likewise,~\cite{salas17} proposes real-time detection of traffic events using tweets and their geolocation information. Finally, \cite{hasan2017survey} offers a survey on event detection in Twitter.

Our proposal \cite{DOMINGUEZ2017} stands out because of two reasons: (i) using a density-based clustering algorithm, it establishes a pattern of the urban area under analysis, which enables further analysis; and (ii) it is able to detect not only unusually big crowds, but also unusually small crowds, crowds in areas that usually have much less activity, and the absence of crowds in areas that usually have higher activity. This proposal offers promising detection results, without the high complexity of other video-based approaches.

\subsection{Cellular-based traces}
\noindent

Mobility at individual level unveils interesting information, since the subsequent locations a person visits define her/him in many ways. Continuously tracking the user and appropriately mining the resulting sequence of locations, indeed disclose much more information that can be used for early detection of abnormal situations in cities (i.e., of incidents).

The first data sets used in the literature were originally collected by surveys. One of the first studies on human mobility not relying on surveys was performed by analyzing the circulation of bank notes in the United States ~\cite{Brockmann2006}. Nevertheless, years later it was demonstrated that the conclusions could hide a correlation of population-based heterogeneity and individual human trajectories due to the nature of collected data.  This situation was completely turned upside-down with the tremendous growth of the use of mobile phones. Nowadays almost every person carries a mobile device, all day long, everywhere he/she goes. Thus, the mobile device is the perfect proxy to track user's mobility.  Mobile devices have demonstrated to be key mobility sensors that allow us to investigate global mobility-related topics such as the habits of people around a city ~\cite{Becker2013}.

Among the several systems integrated in the mobile devices that can be leveraged as location proxies, the most popular one is global positioning system (GPS), as it is the only one providing real location coordinates. Also, Wi-Fi and cellular telephony network can be used to provide location.  There are other systems, like Bluetooth and radio-frequency identification (RFID), which are also used in some specific scenarios to locate people in small size environments.

Focusing on cellular-based data, there is a large number of mobility-related studies handling this type of data sets. However, there are different ways to capture cellular telephony-based location data. The most used approach is to use the data stored in the network nodes themselves, which record the BTS to which the device is connected when the user is making or receiving a call, or sending or receiving a short message, known as "call-detailed records".  These data sets are characterized by their high number of users and long duration. The second methodology to collect this location-related data is recording the cell to which the device is attached every moment. This can be done from the mobile device itself, and thus, the data available is much less extensive. This collection method implies the users installing a dedicated application (e.g., \cite{MobilitApp}) that collects data continuously and sends it to a certain server from time to time.

\mbox{\cite{dobra2015}} analyzes the anonymized CDRs provided by a cell phone service provider in Rwanda, with the provider's network activity for more than three years. They use these traces to develop analytical tools that can identify emergency events in real time. This work is the one that most closely resembles ours, but it seeks incidents at a larger geographical level: at the region or country level, not the city level, and without relying on other sources of information such as social networks. \mbox{\cite{papandrea2016}} and \mbox{\cite{keramat2016}} use CDR traces containing voice calls, SMS and Internet from almost 1 million users of a mobile operator in the city of Milan, to analyze the city's points of interest (PoI) and study how people move through them. \mbox{\cite{leo2016}} analyzes a CDR trace of 8 million users for 12 months in Mexico. Using this data, they show the properties of user movements between calls. \mbox{\cite{silveira2016}} uses CDR traces collected in different major cities in Brazil and Mexico. They combine these CDR traces with information from social networks (georeferenced tweets) to predict human mobility, but not oriented to detect incidents. \mbox{\cite{gramaglia2017}} uses traces of user paths extracted from CDR records published by Orange and the University of Minnesota. They use this data to propose solutions to the problem of preserving the privacy of mobile subscriber paths. \mbox{\cite{hong2019}} uses an 8 months CDR trace from Mexico to study migration behaviors. Finally, \mbox{\cite{mamei2019}} uses a CDR data set from an Italian Telecom operator in an area of about 20 million people, spanning several months, to obtain origin-destination matrices of the population in that area.

\subsection{Vehicular adhoc networks}
\noindent

VANETs can be used both to (i) gather information about the traffic state in the streets of the city or detect any traffic incident, as well as to (ii) disseminate warning messages related to any situation in the city that mught affect drivers. For both goals it is necessary to have an efficient routing protocol able to cope to the inherent dynamic features of VANETs. 

Related to the former, we highlight two multimetric proposals that consider several metrics to select the best forwarding node: (i) I-GPSR (Improvement GPSR) \cite{I-GPSR} that uses four metrics (distance to destination, vehicle density, trajectory and vehicle speed); (ii) and VIRTUS \cite{VIRTUS} that implements a multipath routing protocol for video-streaming over VANETs considering the different type of video frames. Our proposal 3MRP \cite{TVT_Ahmad_2017} also uses four metrics (distance to destination, vehicle density, trajectory and available bandwidth). 3MRP is map-aware considering possible obstacles in the city including buildings. It also considers an additional metric of MAC losses, and also arranges a game-theoretical multipath forwarding path according to the different types of video frames. Finally, metrics are weighted dynamically throughout time so that those most decisive metrics are highlighted. Simulation results show that our proposal outperforms the other protocols in terms of packet losses and average packet delay. Other recent proposals about routing protocols for VANETs also use several metrics to take their forwarding decisions. For instance, in \cite{MM_KAS_19}, authors propose an opportunistic routing protocol with which vehicles use three metrics: packet advancement (its value increases with packet
movement towards destination), vehicle density, and packet delivery probability as parameters for determining the appropriate vehicle to be the next hop.

Regarding the latter, we highlight three approaches to disseminate warning messages efficiently in a VANET: (i) Junction-Store-Forward (JSF) \cite{JSF}  is a protocol designed to exploit the road topology by considering that vehicles rebroadcasts the message every time they arrive at a new junction until the message timer expires; (ii) In~\cite{Sanguesa2014a}, authors designed two approaches for vehicular dissemination: Neighbor store and forward (NFS), an opportunistic protocol for scenarios with low vehicles' density so that the current holding vehicle carries the packet until it finds a new neighbor to rebroadcast the message; and nearest junction located (NJL), a scheme for vehicular scenarios with high density designed to relay the message only if the vehicle is the closest to any intersection; (iii) In \cite{DISS_ULL_19}, authors propose a multi-hop broadcast scheme based on an accurate estimation of the connectivity probability among vehicles to coordinate the contention among candidate forwarders. Our proposal adaptive distributed dissemination (ADD)~\cite{Iza18} improves other proposals as results in section \ref{sec:Sensing_Actuating_UPC} show, in terms of packet delivery ratio, average packet delay, broadcast overhead, and number of collision packets.

\subsection{Data Privacy and Trust}
\noindent

Many research efforts have been put in privacy protection of structured data like relational databases. In this field, authors profit from the structure of data to anonymize attributes that are known to be probable identifiers. However, sanitization of unstructured data (query logs, raw text documents) has been taken less into consideration in the literature, despite being the usual way in which data is transferred between parties. Because raw text does not necessarily include explicit identifiers, document sanitization in this case implies two different tasks: the detection of identifiers in the text, and hiding the information. This second step has to be executed so that the disclosure risk is minimized and the utility of the sanitized text is maximized. This trade-off is present in all studies.
Traditionally, sanitization of text documents was done manually by qualified reviewers. That made the process more expensive and time-consuming, apart from not scalable when the data volume grew. Thus, the need for automatic text sanitization methods emerged. 
One of the first unsupervised approaches was the Scrub system~\cite{sweeney96amia} which focuses on the removal of sensitive terms from medical records, being an example of methods that rely on detection patterns for specific data types. The detected terms are then replaced by others of similar type. The main drawback of patterns related to specific areas of knowledge, as well as trained classifiers, is the restriction in applicability of the approach when sanitization needs to be applied to a heterogeneous scenario.
The ERASE system~\cite{Chakaravarthy08CIKM} was developed to perform unstructured text documents sanitization automatically. This is done not by searching for predefined patterns but by detecting and removing sensitive elements using a database of entities (persons, products, diseases, etc.) which aims to model public knowledge. Each entity in the database is associated with a set of terms, which constitute the context of the entity, and some entities are considered protected. Thus, ERASE works by finding common terms between the document and the entities’ context, and removing them from the document. In this way, no protected entity can be inferred as being mentioned in the document by an adversary. 
The main disadvantage of removing sensitive terms is the decrease in the document’s utility. Moreover, the obscuration may raise awareness of the document’s sensitivity.  
Another approach~\cite{Sanchez13INS} relies on external general-purpose knowledge bases, which broadens the applicability of the method. Because it requires no supervision during the sanitization process, it is a more scalable solution and enables its application to environments with large data volumes. The detection of sensitive terms in this case relies on quantifying how much information each textual term provides. Thus, a term is considered sensitive if it provides more information than what the attacker is assumed to possess. In order to do that, the Information Content (IC) of each textual term is computed. Because the IC of a term is, formally, the inverse of its appearance probability in a corpus, a common approach to determine this probability in the Web corpus is to use the hit count provided by web search engines when querying a term. An improvement of this method was developed by its authors~\cite{Sanchez14INS} enabling the quantification of disclosure risk of semantically correlated terms with a sensitive one whether the latter is removed or generalised. Moreover, the terms that may cause disclosure of sensitive ones are generalized rather than removed, which increases the final document’s utility significantly.

\section{Conclusions and future work}
\label{sec:conclusions}
\noindent

In this work we have presented the INRISCO~\cite{INRISCO} framework whose aim is the early detection of incidents in urban environments. After detecting these unexpected behaviors, INRISCO also performs an efficient smart dissemination of warning messages through vehicular networks and also through networks based on BLE or Wi-Fi. This way, INRISCO can rapidly warn citizens about that incident. 

Regarding the detecting process, INRISCO uses citizens' activity in LBSNs in order to infer typical behavioral patterns in the urban area under study. Those patterns are used as reference to compare with the current (on-the-fly) behavior. INRISCO understands that and incident is happening when significant differences are found between the reference pattern and the current behavior. Those evidences are also combined with other two supplementary mechanisms. On the one hand, by using entropy rate variations as an indicator of unexpected changes in citizens' activity in social media. On the other hand, by analyzing the textual content of the data collected from social media, the INRISCO framework identifies the possible causes behind the anomalies detected from the citizens' movements.

Of course, the information taken from social media has to be checked. Thus, the INRISCO framework allows us filtering shared information in social networks from analyzing two perspectives: (i) credibility of Tweets' content and their metadata, and (ii) authenticity of the sources. The results obtained have been definitively satisfactory. Furthermore, we claim that results could even become better if both perspectives were combined, which we will develop in a future work.

Regarding the dissemination mechanisms, INRISCO simultaneously uses several supplementary ways. On the one hand, the dissemination in VANETs has been tackled in urban scenarios. Our goal is to avoid broadcast storms, which are especially sensitive in cities because of the potential high density of vehicles, numerous crossroads and signal blocking mainly due to buildings. On the other hand, we have also disseminated the alert messages by using the available networks in the surroundings of the detected event or incident, such as BLE and Wi-Fi. By defining specific protocols, INRISCO is able to disseminate messages without overloading the network and without reaching beyond the area of interest (proximity).

To conclude, the INRISCO platform exploits the fact that citizens can be considered as sensors in movement across the city: they carry smart phones, drive in connected vehicles, participate in social networks. Those so-called smart citizens produce a huge amount of information, which properly processed can feed interesting applications for smart communities. This way, the product of the data is returned to the citizens themselves. INRISCO is able to (i) rapidly detect incidents in the city, (ii) analyze the danger level of the incident, and (iii) quickly respond to the incident by alerting emergency units and also by disseminating warning messages to citizens.

\section*{Acronyms}
{\small 
\begin{tabular}{ l l }

3MRP & Multimedia multimetric map-aware routing protocol \\
ABE & Available bandwidth estimation  \\
ADD & Adaptive distributed dissemination  \\
API & Application programming interface  \\
ATB & Adaptive traffic beacon   \\
BLE & Bluetooth low energy    \\
CDR & Call-detail records \\
CTD  &    Collaboratively triggered dissemination    \\
DBSCAN   &  Density-based algorithm
for discovering clusters in large spatial databases with noise  \\
DSW  & Dynamic self-configured weights    \\
DTM  &  Document term matrix    \\
ESD & Early story detection   \\
FDR  & Frame delivery ratio    \\
GPS &  Global positioning system    \\
GPSR  & Greedy perimeter stateless routing    \\
INRISCO  & INcident monitoRing In Smart COmmunities    \\
JSF &  Junction store and forward    \\
KNN & K-nearest neighbor \\
LBSN  & Location-based social networks    \\
LSH  & Locality sensitive hashing \\
LZ  & Lempel Ziv    \\
MANETs  &    Mobile ad hoc networks    \\
MLP & Multilayer perceptron  \\
NSF  & Neighbor store and forward    \\
NYC  & New York city    \\
OSM  & Open street map    \\
PSNR &  Peak signal-to-noise ratio    \\
QoS &  Quality of Service    \\
RC  & Reference clusters    \\
RSU &  Road side unit    \\
SCF &  Store carry forward    \\
SDC  & Statistical disclosure control    \\
SUMO  & Simulation of urban mobility    \\
TDM  &  Term document matrix    \\
UUID  & Universally unique identifier    \\ 
UV-CAST & Urban vehicular broadcast   \\
VANETs   &   Vehicular ad hoc networks    \\
\end{tabular}}
\normalsize



\end{document}